\documentclass[twocolumn]{aastex631}

\usepackage{bm}

\defcitealias{le_active_2024}{Paper-I}
\defcitealias{2026arXiv260110372Q}{Paper-II}

\usepackage{bbding}
\usepackage{lineno}

\shorttitle{}
\shortauthors{Xia et al.}

\begin{document}

\title{Active Galactic Nuclei and STaR fOrmation in Nearby Galaxies (AGNSTRONG). III. A Study on Ionized and {Warm Molecular} Gas Outflows of 6 Type-2 AGNs}

\author[0009-0005-3916-1455]{Ruisong Xia}
\author[0009-0000-0126-8701]{Chen Qin}
\author[0000-0003-1270-9802]{Huynh Anh N. Le\textsuperscript{\Envelope}}
\author[0000-0002-1935-8104]{Yongquan Xue\textsuperscript{\Envelope}}
\author[0000-0002-1653-4969]{Shifu Zhu}
\author[0009-0003-5280-0755]{Mengqiu Huang}
\author[0000-0001-5525-0400]{Hao Liu}
\author[0000-0002-4926-1362]{Xiaozhi Lin}

\affiliation{Department of Astronomy, University of Science and Technology of China, Hefei 230026, China; lha@ustc.edu.cn, 
xuey@ustc.edu.cn}
\affiliation{School of Astronomy and Space Science, University of Science and Technology of China, Hefei 230026, China}

\begin{abstract}
Active galactic nucleus (AGN)-driven gas outflows are one of the best tracers of AGN feedback in action, as these powerful outflows expel/heat or compress the surrounding interstellar medium (ISM), thus quenching or enhancing star-forming activity in their hosts. 
Studying the kinematics of outflows in different gas phases is crucial for comprehending how AGNs impact the ISM within their host galaxies. 
However, the differences in the physical natures of ionized and warm molecular gas outflows remain largely unexplored. To obtain a complete picture of AGN outflows and their feedback effects, we present a study of both ionized and warm molecular gas outflows in six type-2 AGNs ($z<0.1$) that exhibit strong ionized outflows in previous optical observations. 
Utilizing the Triple Spectrograph and Double Spectrograph instruments on the Palomar 200-inch Hale Telescope, we conduct spatially resolved measurements in the slit direction of strong emission lines from both ionized and warm molecular gas, such as $\rm [O\ III]$, $\rm Pa\alpha$, $\rm H_{2}$ 1-0 S(1), etc., allowing for a direct comparison of their outflow properties. 
One out of six AGNs shows significant ionized and warm molecular outflows in near-infrared bands, exhibiting the most powerful kinematics and highest luminosity. A positive correlation between the kinematics and AGN luminosity is shown, suggesting that more luminous AGNs, which reflect higher levels of AGN activity, tend to have a greater impact on the gases, probably driving the outflows.

\end{abstract}


\section{INTRODUCTION} \label{sec:intro}

The correlations between the mass of {the central} supermassive black hole (SMBH) and the properties of the host galaxy hint at a co-evolution of {black holes (BHs)} and galaxies, which may be regulated by {active galactic nucleus (AGN)} feedback (e.g., \citealt{kormendy_coevolution_2013,xue_chandra_2017,caglar_llama_2020}), {as predicted} by theoretical studies. The two primary scenarios, negative and positive feedback, are employed to interpret the observed results {(e.g., \citealt{zubovas_galaxy-wide_2014,le_ionized-gas_2017,le_fluorescent_2017, 2023ApJ...945...59L})}. 
In these scenarios, AGN feedback can either expel/heat or compress the gas in the interstellar medium (ISM), thus suppressing or enhancing star-forming activity in their hosts \citep[e.g.,][]{2015A&A...582A..63C, 2015ApJ...799...82C,2016A&A...591A..28C,2022MNRAS.512L..54B}. 
However, some studies do not reveal clear signatures of feedback.
For instance, studies across various samples have identified flat relationships between the star-formation rate (SFR) and AGN luminosity (e.g., \citealt{harrison_no_2012,stanley_remarkably_2015, schulze_no_2019, ramasawmy_flat_2019}), suggesting a lack of clear evidence for the enhancement or suppression of star formation \citep{ruschel-dutra_agnifs_2021}. These findings underscore the complexity of AGN feedback and the intricate interactions between the nuclear activity and the ISM, challenging simplistic explanations.

{AGN-driven} gas outflows are one of the best tracers of AGN feedback. Over the past two decades, most previous works have focused on outflows traced by optical emission lines (e.g., \citealt{karouzos_unraveling_2016,harrison_agn_2018}). {Studies} on outflows involving different gas phases, such as warm molecular gas {(e.g., rovibrational $\rm H_2$ emission lines) in near-infrared (NIR) bands}, are less common but gaining popularity (e.g., \citealt{storchi-bergmann_feeding_2009,u_inner_2013,rupke_breaking_2013,riffel_molecular_2013,riffel_feeding_2014,izotov_near-infrared_2016,ramosalmeida_near-infrared_2019,riffel_active_2020}). 
Ionized gas and {warm molecular} gas are both important and complementary to each other, and comparative studies between them have garnered significant {interests} in the field of AGN feeding and feedback, finding that different gas phases show different properties when tracing the feeding or feedback. For instance, in an Integral Field Spectroscopy (IFU) study, \citet{storchi-bergmann_feeding_2009} suggested that $\rm H_2$ gas traces the AGN feeding by the different intensity distribution compared with the bi-conic region detected by ionized gas. In contrast, also using IFU observations, \citet{rupke_breaking_2013} revealed a compact disk and {warm molecular} outflow traced by $\rm H_2$. The $\rm H_2$ outflow gas has a bi-conic angle of 100 degrees and extends perpendicular to the disk up to a distance of 400 pc from the {nucleus}. \citet{ramosalmeida_near-infrared_2019} found that outflows of ionized gas are faster than those of {warm molecular} gas. In addition, they also found that {warm molecular} gas {outflows have} a significant impact on the host galaxy, suggesting that outflows traced by {warm} molecular $\rm H_2$ gas may play an important role in regulating the growth of SMBHs and their host galaxies.

The general distinctions between the nature of {warm molecular} gas and ionized gas outflows remain unclear, including differences in size, kinematics, strength, and their impact on the surrounding ISM. These differences are crucial for comprehending the outflows and AGN feedback.  {Warm molecular} gas, being cooler, is typically in the outer regions of the AGN structure. However, previous measurements of outflow sizes in the literature have predominantly relied on ionized gas, which is located in the inner parts of AGNs, potentially resulting in underestimation of {the sizes of the outflows}. Therefore, accurately estimating outflow {sizes} based on the kinematics of {warm molecular} gas is of significant importance. It is thus necessary to continue comparative studies between {warm molecular} gas and ionized gas with more samples.

While optical spectra of AGNs predominantly exhibit strong emission lines from ionized gas \citep[e.g.,][]{karouzos_unraveling_2016, woo_prevalence_2016, bae_limited_2017, ruschel-dutra_agnifs_2021}, NIR spectra of AGNs offer the advantage of revealing emission lines from both ionized and {warm molecular} gas \citep[e.g.,][]{riffel_0824_2006, riffel_gemini_2021, riffel_agnifs_2021, riffel_agnifs_2023, storchi-bergmann_feeding_2009, rupke_breaking_2013, ramosalmeida_near-infrared_2019,   bianchin_gemini_2022}.
{Utilizing the Triple Spectrograph (TPSP; \citealt{herter_performance_2008})}, a NIR slit spectrograph, we can explore {warm molecular} gas outflows through ro-vibration $\rm H_2$ emission lines, as well as ionized {gas} with multiple emission lines such as {$\rm [S\ III]\ 0.9533\ \mu m$}, $\rm Pa\alpha$, and $\rm Pa\beta$, among others. Our goal is to study the outflow kinematics for gas in different phases. By obtaining the radial profile of $\rm H_2$ emission line, we will determine the kinematics of the {warm molecular} outflows and measure the {warm molecular} outflow size. We will compare the sizes of outflows in different gas phases. In addition, we will also derive mass outflow rates and kinetic power of the {warm molecular} outflows and compare them with those of ionized gas outflows.

To facilitate a comprehensive comparison of outflow properties across different wavelength bands, it is essential for our sample to have undergone detailed optical observations. To achieve this, we follow the sample of 6 type-2 AGNs known for their extreme {ionized} outflows as studied by \citet{karouzos_unraveling_2016}. 
\citet{karouzos_unraveling_2016} and \citet{woo_prevalence_2016} carried out comprehensive examinations of outflows of ionized gas in these AGNs using IFU observations.
These studies revealed a ring-like structure in the star-forming region at the edge of the outflow. Notably, these studies indicated an increase in {SFRs} following the increase of outflow strengths. However, a comprehensive exploration of {warm molecular} gas outflows in these specific objects is yet to be undertaken.

As the {third} paper of AGNSTRONG ({Active Galactic Nuclei and STaR fOrmation in Nearby Galaxies;} see \citetalias{le_active_2024}, i.e., \citealt{le_active_2024}), we observed {these} 6 type-2 AGNs with TPSP and Double Spectrograph (DBSP) of the P200 {telescope} and obtained spectra in the UV-optical and NIR {bands}. Our analysis involves both integrated and spatially resolved measurements of strong emission lines from both ionized and {warm molecular} gas. 
{We extract the outflowing components to measure the properties and} compare the outflow size and kinematics between {ionized and {warm molecular} gas}. The paper is organized as follows: In Section~\ref{sec:data}, we outline the sample selection, details of target observations, and data reduction. Section~\ref{sec:analysis} covers the extraction of integrated and spatially {resolved} spectra, as well as the fitting of emission lines using Gaussian models. Gas properties are estimated and the results are {also} presented in Section~\ref{sec:analysis}, which are discussed in Section~\ref{sec:dis}. Finally, we summarize our work in Section~\ref{sec:conc}. All spectra have been converted to the rest-frame {wavelengths}. Throughout this work, we adopt a flat $\rm \Lambda CDM$ cosmology with $\Omega_{\Lambda} = 0.7$, $\Omega_{\rm m} = 0.3$, and $H_0 = 70~{\rm km~s^{-1}~Mpc^{-1}}$.

\section{SAMPLE AND DATA} \label{sec:data}
\subsection{Sample Selection}

\begin{figure}[t]
\plotone{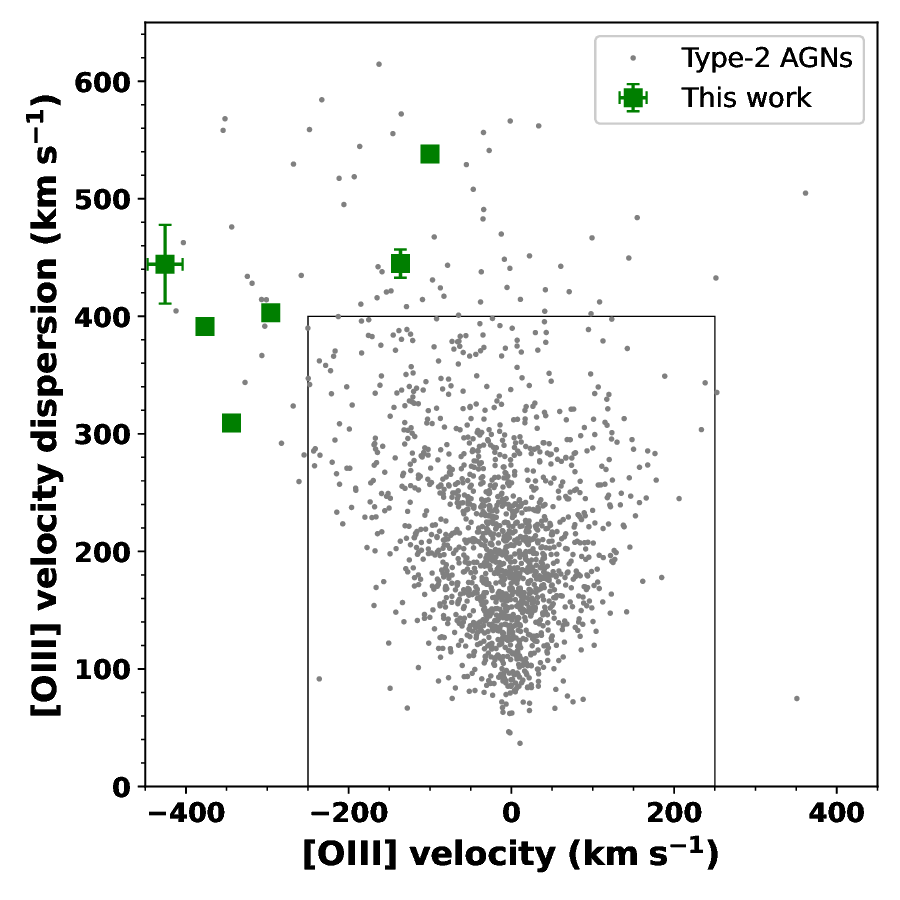}
\caption{[O III] diagram of the luminosity-limited sample of AGNs with {dust-corrected} [O III] luminosity $L_{\rm [O\ III];cor} > 10^{42}\rm\ erg\ s^{-1}$ and $z <0.1$ from \citet{woo_prevalence_2016}. 
We highlight our sample of six {type-2} AGNs using the green squares, while the solid box serves as a reference for {high velocity ($|V|>250 \rm \ km\ s^{-1}$) or large velocity dispersion ($\sigma>400 \rm \ km\ s^{-1}$).
{Figure adapted from \citet{karouzos_unraveling_2016}.}
}
\label{fig:sample}}
\end{figure}

\citet{karouzos_unraveling_2016} conducted a comprehensive analysis of six AGNs with prominent outflow signatures.
They performed a detailed examination of the outflow kinematics of the $\rm [O\ III]$ and $\rm H\alpha$ emission lines of these six AGNs using the Gemini Multi-Object Spectrograph (GMOS) IFU.
However, a comprehensive investigation of {warm molecular} gas outflows in these specific objects has not yet been conducted. 
To delve deeper into the realm of {warm molecular} gas outflows and make comparisons between ionized and {warm molecular} gas properties, we conduct observations using the NIR and optical spectroscopy to detect both ionized and {warm molecular} gas in these six objects. 
{These AGNs are type-2 Seyferts with $L_{\rm [O\,III];cor} > 10^{42}\ \rm erg\ s^{-1}$ and serve as a complementary sample to \citetalias{le_active_2024}, which focuses on type-1 AGNs, thereby providing a more comprehensive view.}
The target names are provided in Table~\ref{tab:log}, and a unique ID is assigned to each target, following an ascending order of right ascension.

{The parent sample of these six AGNs was drawn from \citet{woo_prevalence_2016}. 
By selecting objects with sufficient signal-to-noise ratio (S/N) in both the continuum and emission lines, their sample comprised 38,948 type-2 AGNs. 
From this sample, \citet{karouzos_unraveling_2016} selected 902 AGNs by limiting the redshift to $z < 1$ and requiring high luminosity, with extinction-corrected [O\,III] luminosity $L_{\rm [O\,III];cor} > 10^{42}\ \rm erg\ s^{-1}$. Here, $L_{\rm [O\,III];cor}$ refers to the [O III] luminosity corrected for extinction using the law of \citet{1999Ap&SS.266..243C}, following \citet{bae_census_2014}.
Among these, 29 AGNs were further selected by \citet{karouzos_unraveling_2016} for having high velocity ($|V|>250\ \rm km\ s^{-1}$) or large velocity dispersion ($\sigma>400\ \rm km\ s^{-1}$). 
The six AGNs are part of this subset and were observed with the GMOS IFU \citep{2002PASP..114..892A} on Gemini-North \citep{karouzos_unraveling_2016}.
Their ionized gas exhibits robust outflow signatures, characterized by high velocities or large velocity dispersions, as illustrated in Figure~\ref{fig:sample} after redshift correction.
}

\subsection{TPSP Observations}

\begin{figure*}[!ht]
\centering
\includegraphics[scale=0.53]{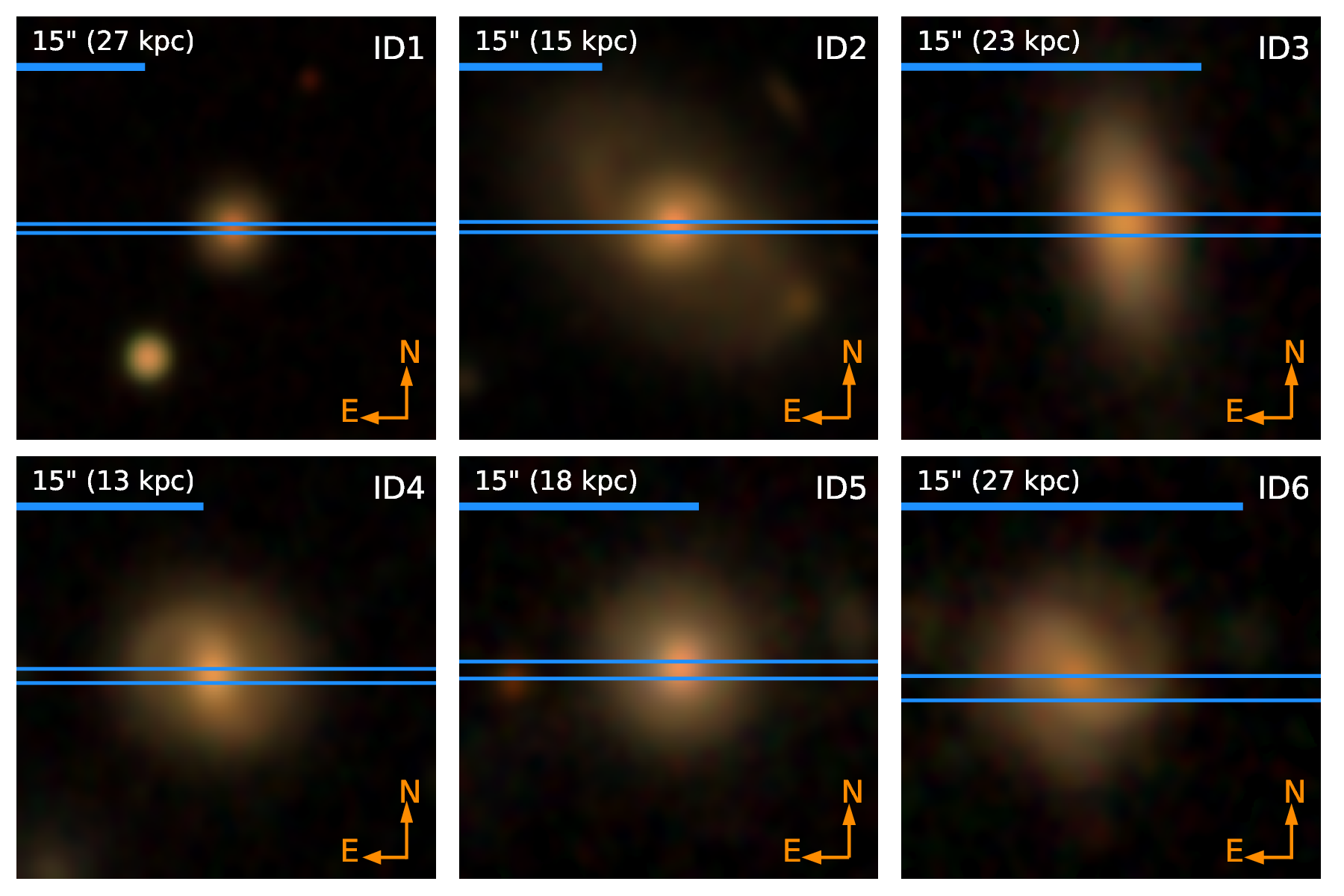}
\caption{SDSS \textit{gri} composite images of the targets. The blue solid lines show the direction of the slits of TPSP. The scale bar is displayed at the upper left corner of each panel, representing the scale of $\rm 15\arcsec$, and the corresponding spatial size is marked in brackets.}
\label{fig:sdss}
\end{figure*}

We conducted observations of these six {type-2} AGNs using TPSP in the 2022A semester (ID: CTAP2022-A0037; PI: Huynh Anh N. Le). The wavelength coverage is 1.0--2.4 $\rm \mu m$. The slit size is $1\arcsec\times 30\arcsec$, with a scale of 0.28 arcsec per pixel along the slit.
{Figure~\ref{fig:sdss} shows the slit overlaid on the SDSS images of the targets.}
The resolution of the instrument is approximately $\rm 50\ km/s$ ($\rm R=\lambda/\Delta\lambda=2500\text{--}2700$).
Our observations were conducted in a nodding pattern ABBA, with the targets positioned differently along the slit in the A and B modes. A-B image differencing was performed to obtain the sky background spectrum for the other frame.
The single exposure times for our target objects varied from $\rm 160\ s$ to $\rm 500\ s$, as shown in Table~\ref{tab:log}.
To correct telluric absorption lines in the NIR spectra of our sample, we observed several A0 V standard star spectra on the same night. 
The seeing conditions during our observations ranged from $2\arcsec$ to $2.5\arcsec$, while the airmass varied between $1.05$ and $1.34$. By analyzing the full width at half-maximum (FWHM) of five standard stars, we {derived} an approximate average seeing of $2.4 \arcsec$ \citep{rose_quantifying_2018, ramosalmeida_near-infrared_2019}. 

\subsection{DBSP Observations}

Three out of six sources in our sample were observed in the optical band by DBSP in the 2022 semester ({ID: CTAP2022-A0024;} PI: Luming Sun \& Yibo Wang), providing the opportunity to study spatial properties with spatially resolved optical slit spectroscopy.
DBSP is a grating spectrometer at {the} P200 Cassegrain focus, using a dichroic (D55 in this work) to split light into blue and red channels and observe simultaneously. {The wavelength coverage is 3000--11000 $\text{\AA}$}. We used 600/4000 grating to detect high-resolution spectra, with the FWHM of the instrument being approximately
$4.2\ (6.0)\ \text{\AA}$ for the blue (red) channel.
The length of the slit is $128 \arcsec$ and the width is $1.5 \arcsec$. The spatial scale of the individual spatial pixels is $0.389\arcsec$ for the blue channel and $0.293\arcsec$ for the red channel. The exposure time is detailed in Table~\ref{tab:log}. The average seeing of DBSP observations is about $1.3\arcsec$.

\subsection{Data Reduction}
As described in \citetalias{le_active_2024}, the data reduction for TPSP and DBSP observations is carried out using PypeIt \citep{prochaska_pypeit_2020}. It is a Python package for semi-automated reduction of spectroscopic data, supporting more than 20 {s}pectrographs, including DBSP and TPSP.
We follow the procedures outlined in the official cookbook for the flux calibration, co-adding of exposures, and telluric correction. The complete 1D NIR and optical spectra can be seen in Appendix~\ref{appendixA}, shown in Figures~\ref{fig:nir-spec} and \ref{fig:opt-spec}, respectively.
For the three AGNs without DBSP observations, we supplement the optical data with Sloan Digital Sky Survey (SDSS) spectra for our sample, which {were} analyzed by \citet{karouzos_unraveling_2016}.

\begin{figure*}[ht!]
\centering
\includegraphics[scale=0.45]{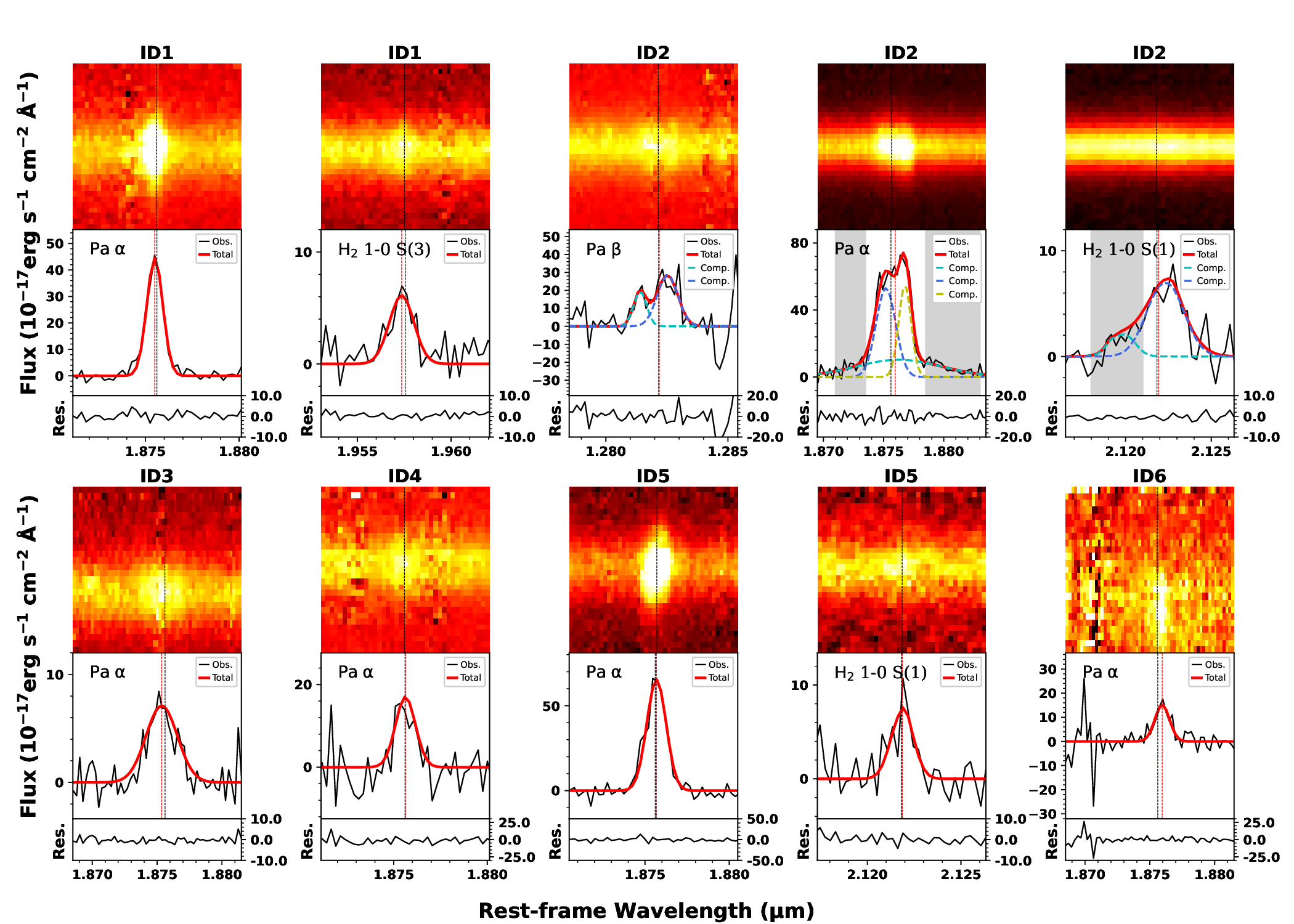}
\caption{Examples of 2D spectral images and integrated emission line profiles for the TPSP observations. The images of 2D spectra close to the emission lines are shown in zscale following the IRAF zscale algorithm (top). 
The x-axis is the wavelength, and the y-axis is the spatial pixels on the slit. 
{The observed integrated emission lines (black) are fitted with Gaussian profiles (middle).
The red curve shows the best-fit model, while the blue, cyan, and yellow curves represent the individual Gaussian components in the double- and triple-Gaussian fits.
The black dashed line represents the reference wavelength of each line, while the red dashed line represents the centroid wavelength of the total line profile.
}
The residuals are shown at the bottom. 
The wavelength ranges used to estimate the outflow size are shaded.
}\label{fig:2dimage}
\end{figure*}

\begin{figure*}[ht!]
\centering
\includegraphics[scale=0.8]{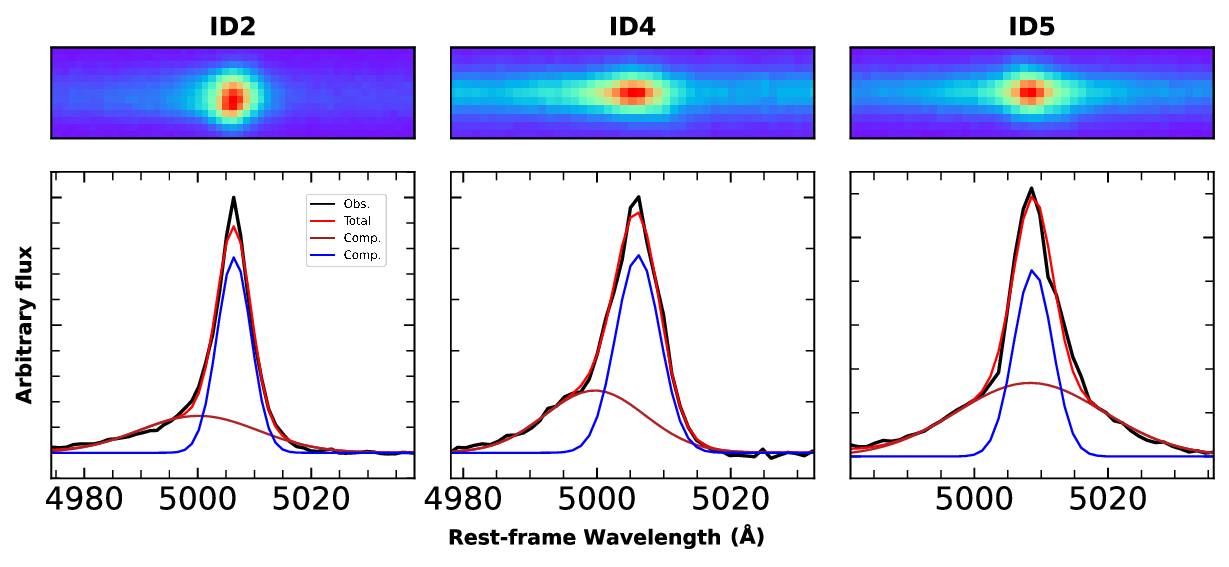}
\caption{{Examples of 2D spectral images and integrated emission line profiles of optical [O III] emission lines in the DBSP observations.
{On the top are the images of 2D spectra.
The observed integrated emission lines (black) are fitted with Gaussian profiles.
The red curve shows the best-fit model, while the blue and brown curves represent the individual Gaussian components.
}
}
}\label{fig:2dimage2}
\end{figure*}

To fully exploit the spatial resolution along the slit to investigate the spatial asymmetry of emission lines, we extract two-dimensional spectra. 
We apply offsets to the \emph{trace} obtained from the 1D reduction process in the two-dimensional spectra to extract a series of 1D spectra.
The \emph{trace} represents the central position of the target on the slit. 
Subsequently, we {follow} the same procedures as the 1D reduction process for each extracted spectrum. 
Examples of 2D spectral images for TPSP and DBSP with the continuum removed are presented at the top of each image of Figures~\ref{fig:2dimage} and \ref{fig:2dimage2}, respectively. The x-axis represents the rest-frame wavelength, while the y-axis corresponds to spatial pixels along the slit. 


\section{MEASUREMENTS AND RESULTS}\label{sec:analysis}  
\subsection{Refining Redshifts}\label{subsec:red}

To better measure the emission line kinematics, we refine the redshifts {estimated by \citet{bae_census_2014}} based on the optical spectra. This involves masking regions containing emission lines with broad components and conducting an initial spectral fit using Penalized Pixel-Fitting (pPXF, \citealt{cappellari_full_2022}). 
pPXF is a versatile method adept at fitting a suite of templates, encompassing both stellar and gas components, to an observed spectrum. Given that all our targets are {type-2} AGNs, we utilized E-MILES simple stellar population models \citep{vazdekis_uv-extended_2016} and single Gaussian emission lines.
pPXF provides estimates for stellar velocity (based on unrefined redshifts) and velocity dispersion based on the best-fit models. The stellar velocity is not negligible, reaching up to tens of $\rm km\ s^{-1}$. The redshifts listed in Table~\ref{tab:log} have been refined using the estimated stellar velocities, which differ by approximately 0.00005 compared to those measured by SDSS.
The spectra utilized in this study have been adjusted according to these refined redshifts.
Therefore, {the [O III] velocity and velocity dispersion of the six selected AGNs (i.e., the green squares in Figure~\ref{fig:sample}) differ} slightly from those in previous works \citep{karouzos_unraveling_2016,woo_prevalence_2016}.

\begin{table*}
 \caption{{Observational} Log} \label{tab:log}
 \centering
 \renewcommand{\arraystretch}{1}
\setlength{\tabcolsep}{1.8mm}{}
\begin{tabular}{cccccccc}
  \hline
  \hline
  ID & SDSS Name & Redshift &$m_{\rm K}$ & $\log\ L_{\rm [O\ III];cor}$ & Instrument & {Observational} Date & Exposure Time \\
  
  & & & (mag)& ($\rm erg\ s^{-1}$) && &(s)\\
  (1)&(2)&(3)&(4)&(5)&(6)&(7)&(8)\\

  \hline
1& {J091807.52+343946.0}  & 0.0973& 13.5 &42.9& TPSP & 2022 Feb 10 &	500$\times$4\\

2& {J113549.08+565708.2}  & 0.0514& 11.4 &43.1& TPSP/DBSP & 2022 Feb 10/2022 Apr 21 &	240$\times$4/500\\

3& {J140452.65+532332.1}  & 0.0813& 13.5 &42.6& TPSP & 2022 Feb 10 &	250$\times$8\\

4& {J160652.16+275539.0}  & 0.0461& 12.7 &42.2& TPSP/DBSP & 2022 Feb 10/2022 Jun 29 &	160$\times$4/600\\

5& {J162232.68+395650.2}  & 0.0631& 13.4 &42.4& TPSP/DBSP & 2022 Feb 10/2022 Jun 29 &	240$\times$4/600\\

6& {J172037.94+294112.4}  & 0.0995& 14.4 &42.3& TPSP & 2022 Feb 10 &	200$\times$8\\
\hline
 \end{tabular}
 \raggedright
 \tablecomments{Columns are as follows: (1) ID {of our objects}; (2) Name of the counterpart in SDSS; (3) Redshift {refined using the estimated stellar velocities}; (4) K-band magnitude; (5) Dust-corrected [O III] luminosity \citep{karouzos_unraveling_2016}; (6) Instrument for observation; (7) {Observational} date; (8) {Total exposure time.}}
\end{table*}

\begin{table*}
\centering
\small
\caption{{NIR Emission line properties and components.}}
\label{tab:lines}
\begin{tabular}{l l l c c c c c c}
\hline \hline
ID & Line & Comp. & Flux & $V$ & $\sigma$ & FWHM & $R_{\rm NLR}$ & S/N \\
 & & & ($10^{-17} \rm erg\,s^{-1} cm^{-2}$) & ($\rm km\,s^{-1}$) & ($\rm km\,s^{-1}$) & ($\rm km\,s^{-1}$) & ($\rm kpc$) & \\
(1)&(2)&(3)&(4)&(5)&(6)&(7)&(8)&(9)\\
\hline
1 & He I 1.0832 & T & $132.1^{+2.9}_{-3.0}$ & $4.5^{+2.6}_{-3.0}$ & $98.5^{+3.3}_{-3.3}$ & $232.1^{+7.9}_{-7.8}$ & $...$ & 6.4 \\
 & $\rm Pa\alpha$ 1.8756 & T & $476.6^{+7.0}_{-6.5}$ & $-17.9^{+1.1}_{-1.1}$ & $47.5^{+1.6}_{-1.8}$ & $111.8^{+3.8}_{-4.3}$ & $0.38^{+0.01}_{-0.01}$ & 34.3 \\
 & $\rm H_{2}$ 1-0 S(3) 1.9576 & T & $101.4^{+7.0}_{-7.3}$ & $-29.7^{+8.0}_{-7.9}$ & $87.6^{+12.2}_{-11.3}$ & $206.3^{+28.8}_{-26.7}$ & $0.60^{+0.07}_{-0.07}$ & 9.2 \\
\hline
2 & [S III] 0.9533 & T & $2048.4^{+17.3}_{-18.3}$ & $69.4^{+2.1}_{-2.2}$ & $196.2^{+2.4}_{-2.2}$ & $462.1^{+5.8}_{-5.2}$ & $0.25^{+0.01}_{-0.01}$ & 27.8 \\
 & He I 1.0832 & T & $1658.9^{+12.7}_{-12.6}$ & $-34.2^{+2.7}_{-3.0}$ & $279.9^{+3.6}_{-3.6}$ & $487.5^{+8.5}_{-8.5}$ & $...$ & 28.4 \\
 & & 1 & $1321.1^{+26.9}_{-37.0}$ & $-55.7^{+5.2}_{-5.9}$ & $305.0^{+5.8}_{-5.3}$ & $718.4^{+13.6}_{-12.4}$ & ... & 22.1 \\
 & & 2 & $338.2^{+40.5}_{-28.2}$ & $49.2^{+3.2}_{-3.8}$ & $104.5^{+8.5}_{-6.4}$ & $246.1^{+19.9}_{-15.1}$ & ... & 9.3 \\
 & [Fe II] 1.2570 & T & $241.2^{+6.9}_{-6.7}$ & $9.7^{+5.0}_{-4.6}$ & $165.1^{+3.1}_{-2.5}$ & $472.7^{+9.8}_{-12.3}$ & $...$ & 8.9 \\
 & & 1 & $143.7^{+13.1}_{-13.2}$ & $-103.1^{+19.7}_{-17.7}$ & $108.4^{+14.6}_{-14.8}$ & $255.2^{+34.4}_{-34.8}$ & ... & 5.9 \\
 & & 2 & $97.5^{+14.3}_{-13.8}$ & $177.3^{+11.3}_{-13.3}$ & $53.2^{+11.9}_{-10.5}$ & $125.3^{+28.0}_{-24.6}$ & ... & 5.1 \\
 & $\rm Pa\beta$ 1.2822 & T & $439.0^{+97.1}_{-14.4}$ & $8.0^{+3.4}_{-3.1}$ & $141.4^{+2.8}_{-2.5}$ & $409.6^{+7.3}_{-409.6}$ & $0.06^{+0.04}_{-0.04}$ & 11.0 \\
 & & 1 & $128.7^{+296.2}_{-7.9}$ & $-176.8^{+142.0}_{-4.0}$ & $36.5^{+95.9}_{-5.4}$ & $86.1^{+226.0}_{-12.8}$ & ... & 4.7 \\
 & & 2 & $302.7^{+16.5}_{-184.1}$ & $85.8^{+90.8}_{-6.9}$ & $85.0^{+4.7}_{-85.0}$ & $200.1^{+11.0}_{-200.1}$ & ... & 8.9 \\
 & [Fe II] 1.6440 & T & $244.8^{+8.2}_{-7.9}$ & $21.6^{+7.8}_{-7.6}$ & $193.4^{+6.1}_{-5.8}$ & $455.4^{+14.4}_{-13.6}$ & $...$ & 10.4 \\
 & $\rm Pa\alpha$ 1.8756 & T & $2720.3^{+35.6}_{-31.0}$ & $57.8^{+7.0}_{-8.1}$ & $359.1^{+19.8}_{-16.7}$ & $486.8^{+3.3}_{-4.9}$ & $0.17^{+0.00}_{-0.00}$ & 41.2 \\
 & & 1* & $927.2^{+35.1}_{-39.2}$ & $105.7^{+23.7}_{-25.5}$ & $\mathbf{570.3^{+38.0}_{-30.4}}$ & $1343.0^{+89.5}_{-71.7}$ & ... & 11.1 \\
 & & 2 & $1083.5^{+56.0}_{-57.3}$ & $-66.9^{+7.5}_{-8.2}$ & $120.3^{+6.0}_{-5.9}$ & $283.3^{+14.1}_{-14.0}$ & ... & 26.9 \\
 & & 3 & $711.9^{+52.0}_{-50.4}$ & $185.4^{+4.2}_{-4.5}$ & $66.9^{+3.9}_{-3.8}$ & $157.6^{+9.3}_{-9.0}$ & ... & 22.9 \\
 & $\rm H_{2}$ 1-0 S(1) 2.1218 & T & $245.8^{+19.2}_{-17.2}$ & $11.6^{+19.5}_{-19.6}$ & $197.5^{+22.3}_{-16.0}$ & $350.2^{+139.6}_{-89.2}$ & $0.25^{+0.01}_{-0.01}$ & 13.9 \\
 & & 1* & $44.0^{+152.3}_{-20.8}$ & $\mathbf{-288.6^{+277.9}_{-79.0}}$ & $81.6^{+127.0}_{-81.6}$ & $192.1^{+299.1}_{-192.1}$ & ... & 3.1 \\
 & & 2 & $197.5^{+30.3}_{-146.8}$ & $73.3^{+57.2}_{-21.5}$ & $135.4^{+38.7}_{-108.1}$ & $318.8^{+91.1}_{-254.5}$ & ... & 12.3 \\
 & $\rm Br\gamma$ 2.1661 & T & $193.8^{+20.3}_{-20.0}$ & $14.8^{+19.9}_{-21.8}$ & $168.3^{+22.1}_{-20.5}$ & $396.4^{+52.1}_{-48.4}$ & $0.19^{+0.08}_{-0.08}$ & 9.3 \\
\hline
3 & [S III] 0.9533 & T & $429.8^{+11.3}_{-10.4}$ & $-150.2^{+8.5}_{-7.6}$ & $263.1^{+7.5}_{-7.8}$ & $619.7^{+17.6}_{-18.3}$ & $...$ & 10.1 \\
 & He I 1.0832 & T & $180.7^{+4.3}_{-4.5}$ & $1.4^{+6.8}_{-6.0}$ & $215.4^{+5.4}_{-5.0}$ & $507.2^{+12.7}_{-11.9}$ & ... & 6.8 \\
 & $\rm Pa\alpha$ 1.8756 & T & $216.2^{+6.5}_{-6.1}$ & $-41.1^{+5.9}_{-5.9}$ & $187.8^{+7.2}_{-6.2}$ & $442.3^{+17.0}_{-14.7}$ & $0.55^{+0.07}_{-0.07}$ & 16.3 \\
\hline
4 & He I 1.8702 & T & $214.4^{+41.4}_{-53.4}$ & $17.6^{+28.1}_{-13.2}$ & $31.9^{+10.3}_{-31.9}$ & $75.2^{+24.4}_{-75.2}$ & $...$ & 3.7 \\
 & $\rm Pa\alpha$ 1.8756 & T & $228.7^{+20.6}_{-19.9}$ & $3.2^{+9.0}_{-8.6}$ & $69.9^{+10.8}_{-10.6}$ & $164.7^{+25.4}_{-24.9}$ & $...$ & 7.2 \\
\hline
5 & [S III] 0.9533 & T & $1271.6^{+20.6}_{-20.8}$ & $5.6^{+2.4}_{-2.4}$ & $108.1^{+2.1}_{-2.1}$ & $254.5^{+4.9}_{-4.8}$ & $0.18^{+0.29}_{-0.29}$ & 12.0 \\
 & He I 1.0832 & T & $848.9^{+14.3}_{-15.4}$ & $64.4^{+3.5}_{-3.7}$ & $186.5^{+4.5}_{-4.9}$ & $439.2^{+10.6}_{-11.7}$ & $0.15^{+0.14}_{-0.14}$ & 10.3 \\
 & $\rm Pa\gamma$ 1.0941 & T & $154.3^{+8.4}_{-7.6}$ & $9.1^{+6.0}_{-5.9}$ & $66.7^{+3.8}_{-4.2}$ & $157.0^{+8.9}_{-10.0}$ & $...$ & 4.0 \\
 & He II 1.2304 & T & $307.3^{+27.0}_{-28.5}$ & $97.2^{+3.7}_{-4.2}$ & $8.8^{+7.6}_{-8.8}$ & $20.8^{+18.0}_{-20.8}$ & $...$ & 9.9 \\
 & [Fe II] 1.2570 & T & $289.6^{+8.0}_{-7.8}$ & $-21.3^{+3.2}_{-3.2}$ & $136.5^{+8.5}_{-7.7}$ & $321.5^{+19.9}_{-18.1}$ & $0.66^{+0.12}_{-0.12}$ & 8.7 \\
 & [Fe II] 1.6440 & T & $214.8^{+3.8}_{-3.8}$ & $-23.9^{+2.3}_{-2.3}$ & $97.6^{+1.9}_{-2.1}$ & $229.8^{+4.6}_{-4.9}$ & $...$ & 9.7 \\
 & $\rm Pa\alpha$ 1.8756 & T & $939.0^{+9.3}_{-8.3}$ & $10.8^{+1.0}_{-1.0}$ & $77.5^{+1.2}_{-1.4}$ & $182.6^{+2.8}_{-3.3}$ & $0.13^{+0.04}_{-0.04}$ & 27.7 \\
 & $\rm H_{2}$ 1-0 S(1) 2.1218 & T & $110.8^{+21.3}_{-20.4}$ & $6.0^{+19.0}_{-19.0}$ & $66.6^{+24.4}_{-34.7}$ & $156.9^{+57.4}_{-81.7}$ & $...$ & 6.4 \\
\hline
6 & $\rm Pa\alpha$ 1.8756 & T & $214.6^{+14.0}_{-11.6}$ & $54.9^{+6.0}_{-6.5}$ & $76.3^{+8.4}_{-9.0}$ & $179.8^{+19.7}_{-21.1}$ & $...$ & 5.1 \\
\hline
\end{tabular}
\begin{raggedright}
\raggedright
\tablecomments{{Columns are as follows: 
(1) ID of our objects; 
(2) the name and wavelength in $\mu m$ of the line; 
(3) the components of the model, with T indicating the total profile and 1, 2, and 3 denoting the individual components;
(4) the flux of the model profile;
(5) the centroid velocity of the profile;
(6) the velocity dispersion of the profile;
(7) the FWHM of the profile;
(8) the radius of the NLR size estimated from the total profile, and the dots imply that the lines or components are not spatially resolved;
(9) the signal-to-noise ratio of the continuum-free emission lines.
Bold numbers denote components with high velocities ($|V| > 250~\mathrm{km\,s^{-1}}$) or large velocity dispersions ($\sigma > 400~\mathrm{km\,s^{-1}}$), which are thought to be signatures of significant outflows.
The corresponding component numbers are marked with an asterisk.}}
\end{raggedright}
\end{table*}

\subsection{Emission-line Fitting}\label{subsec:fitting}

We utilize Bayesian AGN Decomposition Analysis for SDSS Spectra (BADASS; \citealt{sexton_bayesian_2020}) on the SDSS and DBSP spectra. BADASS fits various spectral components simultaneously, including the power-law continuum, stellar emission, and both the narrow and broad components of AGN emission lines.
The fitting process is executed using Markov chain Monte Carlo techniques to provide robust uncertainties of the {model} parameters.

To remove the continuum contribution from the NIR spectra, we fit the spectral region adjacent to the emission line with a low-order polynomial while masking the emission line. The order of polynomials depends on the flatness of the spectra, typically requiring a first-degree polynomial in most instances.
Subsequently, we utilize {\emph{scipy.optimize.curve\_fit}} \citep{2020SciPy-NMeth} to model the emission line profiles with one, two, or three Gaussian components as needed. 
{As prior constraints, we specify the initial guesses as well as the lower and upper bounds for the Gaussian parameters (i.e., amplitude, central wavelength, and standard deviation) to ensure computational efficiency. In general, when a secondary component is indicated, we perform a double-Gaussian fit. However, if the S/N of the additional component is less than 3, we regard it as noise and discard it. In such cases, a single-Gaussian model is preferred. An analogous criterion is applied when a third component is introduced.
We utilize the following formula to compute the S/N of the emission lines (or their components) \citep{le_exposure_2015}:
\begin{equation}
{\rm S/N} = \frac{\sum\limits_i (f_{{\rm L},i}-\bar{f_{\rm C})}}{\sigma(f_{{\rm C},i})\sqrt{n_{\rm L}(1+1/n_{\rm C})}} , 
\end{equation}
where $\sum\limits_i (f_{{\rm L},i}-\bar{f_{\rm C})}$ is the intensity of the line (or their components) after continuum subtraction and $\sigma(f_{{\rm C},i})$ denotes the standard deviation of the continuum. $\sqrt{n_{\rm L}(1+1/n_{\rm C})}$ serves as a statistical correction term, where $n_{\rm L}$ and $n_{\rm C}$ represent the numbers of data points for the line and continuum, respectively.}
Illustrative examples of the integrated emission-line fittings and their residuals are displayed at the bottom of each panel of Figures~\ref{fig:2dimage} and \ref{fig:2dimage2}.

Further analyses are based on the Gaussian profiles of the emission lines (see \citetalias{le_active_2024}). {The velocity and velocity dispersion of the gas are measured using the 1st ($\lambda_0$) and 2nd moments ($\sigma$) of the profile:}
\begin{equation}
\lambda_0 = \frac{\Sigma \lambda f_\lambda}{\Sigma f_\lambda},
\end{equation}
{and}
\begin{equation}
\sigma_\lambda^2 = \frac{\Sigma (\lambda-\lambda_0)^2 f_\lambda}{\Sigma f_\lambda}.
\end{equation}
{And the velocity and velocity dispersion can be written as}
\begin{equation}
V = c\,\frac{\lambda_0 - \lambda_{\rm rest}}{\lambda_{\rm rest}},
\end{equation}
and
\begin{equation}
\sigma = c\,\frac{\sigma_\lambda}{\lambda_{\rm rest}},
\end{equation}
{where $c$ is the speed of light and $\lambda_{\rm rest}$ is the rest-frame wavelength of the emission lines, derived using the redshift corrected for the systemic velocity as described in Section~\ref{subsec:red}.
}
{
The [O III] velocity and velocity dispersion in our sources are shown as green squares in Figure~\ref{fig:sample}.
The results of the NIR spectral fitting are summarized in Table~\ref{tab:lines}.
We detect $\rm Pa\alpha\;1.8756$ $\mu$m in all 6 targets, $\rm He\,I\;1.0832$ $\mu$m in 4 targets, and $\rm [S\,III]\;0.9533$ $\mu$m in 3 targets. 
The $\rm [Fe\,II]$ $1.2570$ $\mu$m, $\rm [Fe\,II]$ $1.6440$ $\mu$m, $\rm H_{2}$ 1-0 S(3) and $\rm H_{2}$ 1-0 S(1) are detected in 2 targets each. 
However, in ID 2, $\rm H_{2}$ 1-0 S(3) is blended with [S VI] 1.9630, and the parameters for these two lines are not listed in Table \ref{tab:lines}. 
The remaining lines ($\rm Pa\beta\ 1.2822$ $\mu$m, $\rm Pa\gamma\ 1.0941$ $\mu$m, $\rm Br\gamma\ 2.1661$ $\mu$m, $\rm He\ II\ 1.2304$ $\mu$m, and $\rm He\ I\ 1.8702$ $\mu$m) are each detected in only one source.
For ID~1, 3, 4, 5, and 6, all emission lines are adequately fitted with a single-Gaussian model. In ID~2, four lines (He I 1.0832 $\mu$m, [Fe II] 1.2570 $\mu$m, Pa$\beta$, H$_2$ 1-0 S(1)) require a double-Gaussian model, while one line (Pa$\alpha$) is fitted with a triple-Gaussian model. 
In all cases where a multi-Gaussian fitting is adopted, the individual components are reported explicitly in Table~\ref{tab:lines}.
Bold numbers denote components with high velocities ($|V| > 250~\mathrm{km\,s^{-1}}$) or large velocity dispersions ($\sigma > 400~\mathrm{km\,s^{-1}}$), which are adopted as reference criteria for identifying components associated with significant outflows.
Therefore, signatures of outflows (both ionized and molecular) are found only in ID~2 in the NIR observations.
However, as discussed by \citet{2011ApJ...737...71Z, 2023ApJ...945...59L}, decomposing emission-line profiles into multiple Gaussian components can be artificial.
Therefore, the identification of significant outflow components should be regarded as indicative only and interpreted with caution.
}
 
The bootstrap method, noted for its versatility \citep{rice_mathematical_2007}, serves as our approach to estimating uncertainties. The spectral errors estimated by PypeIt are assumed to follow a normal distribution with a standard deviation corresponding to the errors at each wavelength. Subsequently, we resample the spectra based on the distributions, conducting {1000} iterations of fitting the new spectra. 
We calculate the 16th and 84th percentiles of the obtained parameters to estimate their $1\sigma$ uncertainties, with the median serving as the reliable result. {Note that the sum of the medians is not strictly equal to the median of the sum. This explains the slight deviation between the summed components and the total measurements in Table~\ref{tab:lines}.}

\subsection{Spatially Resolved Kinematics}\label{subsec:srk}

TPSP provides us with spatially-resolved measurements over several kilo-parsec scales along the slit direction, enabling us to examine the kinematics of both the ionized and {warm molecular} gas.
Empirically, spectra extracted from a single pixel may not have sufficient strength to discern the line profile due to noise for TPSP observations. Therefore, we employ a 3-pixel binning method ($0.28\arcsec /\text{pixel}$, about 1/3 of the average seeing) and measure the binned line profiles as described in {Section}~\ref{subsec:fitting}. 
Note that, as an empirical choice, using the binning size of about 1/3 of the seeing, whether in images or spectra, is generally considered to help avoid oversampling without significantly sacrificing resolution \citep{kaastra_optimal_2016}.

\begin{figure*}
    \centering
    \includegraphics[scale=0.43]{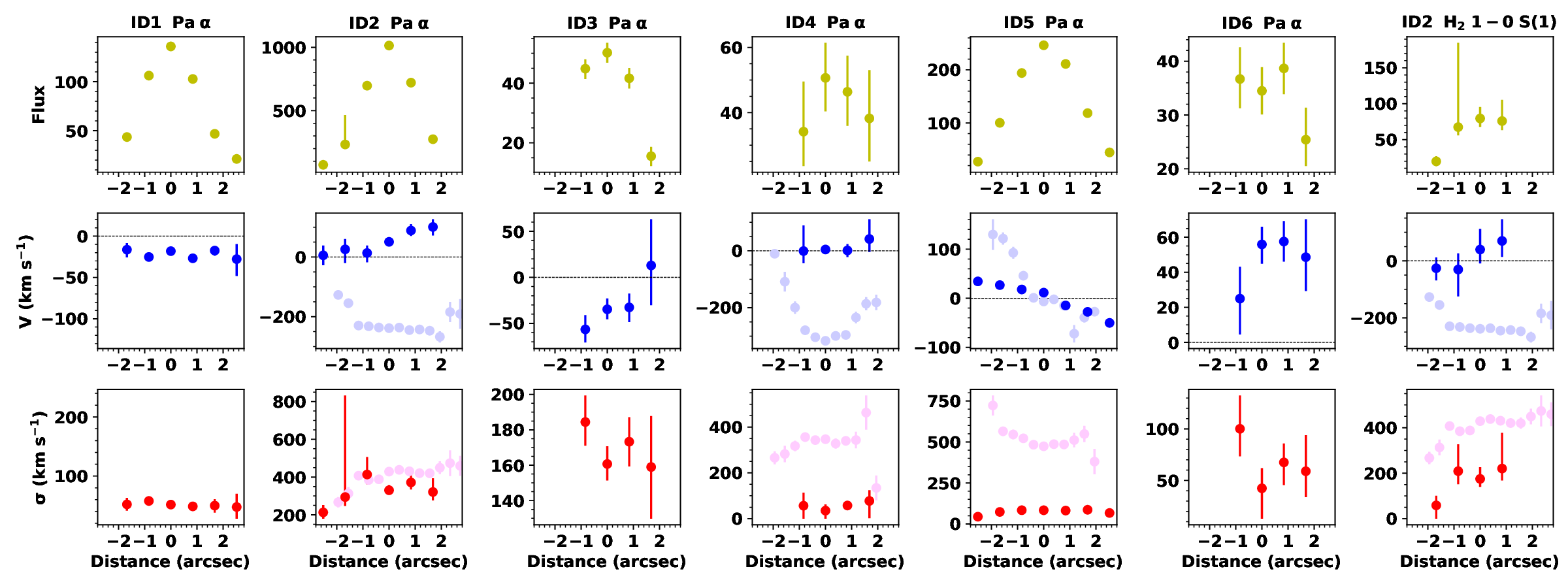}
    \caption{Distributions of flux (yellow dots), velocity (blue), and velocity dispersion (red) of emission lines along the slits. The NIR data are binned with a size of 3 pixels, and the kinematics of optical $\rm [O\ III]$ lines {is} shown in light blue and magenta without binning.}
    \label{fig:distance}
\end{figure*}

{We investigate how the kinematics of the ionized and warm molecular gas vary as a function of distance from the nucleus to the outer regions of the AGN. 
For this purpose, we use the $\rm Pa\alpha$ and H$_2$ 1-0 S(1) lines to represent the ionized and warm molecular gas, respectively, as shown in Figure~\ref{fig:distance}.
Note that though we observed H$_2$ 1-0 S(1) line in ID 5, we found it did not exhibit sufficient spatial resolution. 
Among our sample, only source ID~2 shows clear outflow signatures in the NIR band.
This source also displays enhanced velocity dispersion and an irregular spatial distribution.
The gas kinematics in this source is therefore primarily dominated by outflows.}

\begin{figure*}
    \centering
    \includegraphics[scale=0.43]{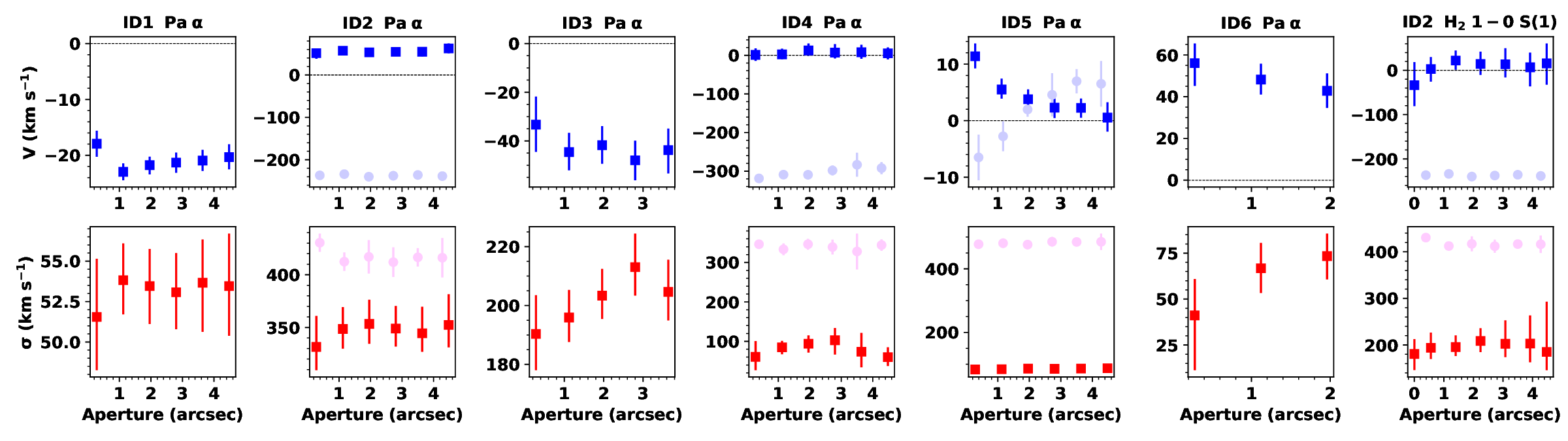}
    \caption{Velocity (blue) and velocity dispersion (red) of NIR emission lines as a function of aperture diameter. The kinematics of optical $\rm [O\ III]$ lines {is} shown in light blue and magenta.}
    \label{fig:aperture}
\end{figure*}

For sources with DBSP observations, the kinematics of [O III] {is} shown in lighter colors for comparison in both Figures~\ref{fig:distance}~and~\ref{fig:aperture}.


Similar to the analysis of kinematics distribution, we present the velocity and velocity dispersion of the emission lines as a function of aperture diameter {({see} Figure~\ref{fig:aperture})}. We examine the kinematic disparity between the inner and outer regions of each galaxy using a range of aperture sizes, spanning from $\sim 1\arcsec$ to $\sim 5\arcsec$. We see no significant trends in the velocity dispersion toward the outer regions.
{Note that in both Figures~\ref{fig:distance} and \ref{fig:aperture}, the velocity and velocity dispersion correspond to the total line profile, since the spatially resolved spectra cannot be consistently and reliably fitted with multi-Gaussian models.
Nevertheless, it is clear that the velocity dispersion of source ID~2 is significantly higher than that of the other sources, with the peak $\sigma$ exceeding that of the others by more than $100\ \mathrm{km\ s^{-1}}$.}

\subsection{Narrow-line Region Size} \label{subsec:ext}

We present examples of NIR 2D spectra in Figure~\ref{fig:2dimage}, most of which exhibit features of spatial extension. Asymmetry also appears in some images, {e.g.,} $\rm Pa\alpha$ of ID~2. The blue-shifted component exhibits a displacement towards the north in the image, whereas the red-shifted component displays a displacement towards the south. 

To fully exploit the spatial resolution potential of TPSP, we investigate the radius of the narrow-line region (NLR) size with emission lines that are sufficiently strong. We sum up the flux in the region close to the emission lines with the continuum removed for each spatial slice along the slit and fit the spatial profiles with a single Gaussian. 
Given the varying seeing conditions and redshifts of our target sources, not all of our {sources} are spatially resolved. Following the criteria outlined by \citet{rose_quantifying_2018} and \citet{ramosalmeida_near-infrared_2019}, we consider targets to be spatially resolved when the FWHM measurements taken from emission lines are more than $3\sigma$ larger than the FWHM measured in the same region from the standard stars.
The radius of NLR ($R_{\rm NLR}$) on the slit can be calculated {as}
\begin{equation}\label{eq:NLRsize}
    R_{\rm NLR} = \rm \frac12\sqrt{FWHM_{obs}^2 - FWHM_{seeing}^2}\ ,
\end{equation}
where the $\rm FWHM_{seeing}$ {is} carefully measured {using} the same method as $\rm FWHM_{obs}$, with the spectra of standard stars in the wavelength range close to the emission line. 
When the size of NLR notably surpasses the seeing \citep{rose_quantifying_2018}, it is considered to {be resolved}:
\begin{equation}
    \rm FWHM_{obs} \geq FWHM_{seeing} + 3\sigma\ ,
\end{equation}
where $\sigma$ is the uncertainty of the seeing.
The results are shown in column~(8) of Table~\ref{tab:lines}. 
Note that the spatial profiles may be partially obscured by noise, which could lead to an underestimation of the sizes.

\subsection{Mass of Ionized and {Warm Molecular} Gas}

Utilizing the fluxes of H$_2$ 1-0 S(1), $\rm Pa\alpha$, $\rm Br\gamma$, and $\rm Pa\beta$ emission lines, it is possible to derive the masses of the {warm molecular} gas and the ionized gas. Specifically, the mass of the {warm molecular} gas can be estimated by \citep{scoville_velocity_1982,riffel_feeding_2014, riffel_agnifs_2023, bianchin_gemini_2022}
\begin{equation}\label{eq:MH2}
    (\frac{M_{\rm H_2}}{M_\odot})=5.0776\times 10^{13}(\frac{F_{\rm H_2\ 1-0\ S(1)}}{\rm erg\;s^{-1}\;cm^{-2}})(\frac{D}{\rm Mpc})^2,
\end{equation}
where $F_{\rm H_2\ 1-0\ S(1)}$ represents the flux of the H$_2$ 1-0 S(1) emission line, and $D$ denotes the luminosity distance calculated using the redshift of the galaxy, assuming thermodynamic equilibrium with an excitation temperature of $\rm 2000\ K$.
In our sample, two H$_2$ 1-0 S(1) lines are detected. And consequently, the mass of {warm molecular} gas for ID 2 is estimated to be {$\log(M_{\rm gas}/M_\odot)=3.82_{-0.03}^{+0.03}$, while for ID 5, it is $\log(M_{\rm gas}/M_\odot)=3.65 ^{+0.08}_{-0.09}$.}

{The mass of the ionized gas can be estimated using \citep{2018MNRAS.474..128R}:
\begin{equation}\label{eq:Mio}
M = \frac{L({\rm H}\beta)m_p}{\alpha^{\rm eff}_{{\rm H}\beta} h \nu_{{\rm H}\beta} n_{\rm e}},
\end{equation}
where $m_p$ is the proton mass,
$h$ is the Planck constant,
$\nu_{{\rm H}\beta}$ is the rest-frame frequency of the H$\beta$ transition,
and $n_{\rm e}$ is the electron density of the gas.
$\alpha^{\rm eff}_{{\rm H}\beta}$ denotes the effective Case B recombination coefficient for H$\beta$ assuming electron temperature $T_e \approx 10^4$~K (consistent with the temperature estimated for IDs 2 and 5).
Finally, $L({\rm H}\beta)$ denotes the H$\beta$ luminosity, which is inferred from the detected emission lines using the theoretical Case~B ratios, e.g., $F({\rm Pa}\alpha)/F({\rm H}\beta) \approx 0.332$, $F({\rm Pa}\beta)/F({\rm H}\beta) \approx 0.163$, and $F({\rm Br}\gamma)/F({\rm H}\beta) \approx 0.028$ \citep{1995MNRAS.272...41S, 2006agna.book.....O}.
The distances are calculated from the redshifts listed in Table~\ref{tab:log}.
Since $\rm Pa\alpha$ is detected in all targets and the electron densities $n_{\rm e}$ are derived from the [S II] doublet for each source (see Appendix~\ref{appendixA}), we estimate the ionized gas masses to be $\log(M_{\rm gas}/M_\odot)=6.28^{+0.01}_{-0.01}$, $6.90^{+0.01}_{-0.01}$, $7.39^{+0.01}_{-0.01}$, $6.45^{+0.04}_{-0.04}$, $6.89^{+0.00}_{-0.00}$, and $7.42^{+0.03}_{-0.02}$ for IDs~1--6, respectively.}

{Similarly, since the $\rm Pa\beta$ and $\rm Br\gamma$ lines are detected in ID~2, we additionally estimate the ionized gas mass from these lines to be $\log(M_{\rm gas}/M_\odot)=6.43^{+0.09}_{-0.01}$ and $6.85^{+0.04}_{-0.05}$, respectively. 
The estimate based on $\rm Br\gamma$ is consistent with that derived from $\rm Pa\alpha$, whereas the estimate based on $\rm Pa\beta$ is lower. This deviation may be caused by telluric absorption, as the $\rm Pa\beta$ line lies close to a wavelength range that is strongly affected by sky emission (see Figure~\ref{fig:nir-spec}). 
Therefore, we adopt the ionized gas mass derived from $\rm Pa\alpha$ as our fiducial estimate for ID~2.
}

\subsection{Outflow Properties}
\begin{figure*}
    \centering
    \includegraphics[scale=0.46]{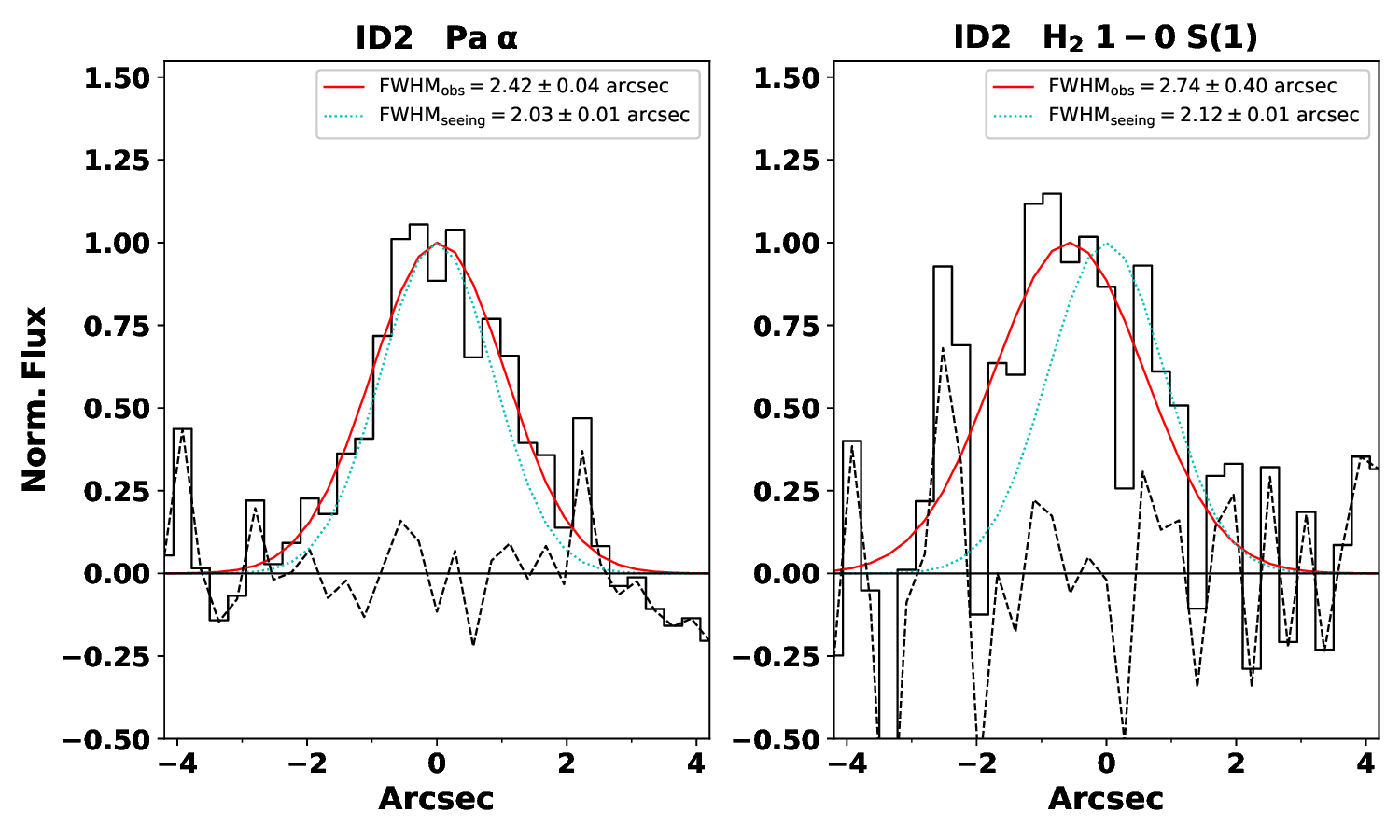}
    \caption{Spatial profiles of the continuum-subtracted outflow components for $\rm Pa\alpha$  and $\rm H_2\ 1\text{-}0\ S(1)$. The fluxes have been normalized by the peak values of the fitted Gaussian profiles, which are represented by the red lines. The residuals are indicated by the dotted lines and the cyan dotted lines correspond to the seeing FWHM.}
    \label{fig:rout}
\end{figure*}

We determine the wavelength ranges used to estimate the outflow size following the same procedure described in Section~\ref{subsec:ext}. The wavelength ranges are shaded in Figure~\ref{fig:2dimage} for lines with {significant outflow} components, {as marked with an asterisk in Table~\ref{tab:lines}}. 
The spatial profiles of the continuum-subtracted {outflow} components for these lines ($\rm Pa\alpha$, and $\rm H_2\ 1\text{-}0\ S(1)$) are shown in Figure~\ref{fig:rout}. 
We see that the profile of $\rm H_2$ is skewed to the left, indicating that the intensity distribution of the {warm molecular} outflow is different from that of the ionized outflow in ID 2.
Similar to Equation~(\ref{eq:NLRsize}), the outflow size is calculated by
\begin{equation}
    R_{\rm out} = \rm \frac12\sqrt{FWHM_{obs}^2 - FWHM_{seeing}^2}\ .
\end{equation}
The outflow sizes estimated from these two lines are $0.66 ^{+0.02}_{-0.02} \rm kpc$ and $0.87 ^{+0.20}_{-0.20} \rm kpc$, respectively. {Compared to the outflow size estimated from the ionized emission lines, the size inferred from the warm molecular lines is larger, albeit with a broad range of uncertainties.}

{Assuming that the outflow is traced with component with high velocities or large velocity dispersions, we use the fluxes of the outflow H$_2$ 1-0 S(1) and Pa$\alpha$ components to calculate the outflow masses via Equation~(\ref{eq:Mio}), adopting the same electron densities as those used for the NLR (see Appendix~\ref{appendixA}).}

Assuming a biconical outflow model \citep{bae_prevalence_2016}, we estimate the {maximum outflow velocity by adding velocity dispersion $\sigma$ to the systemic velocity $V$} \citep{2023ApJ...951....7A}:
\begin{equation}
    V' = 2 \sqrt{V^2 + \sigma^2}.
\end{equation}

The outflow rates of mass and energy are calculated {as} \citep{2017ApJ...836...11G,2023ApJ...951....7A}
\begin{equation}
    \dot{M}_{\rm out} = 3(\frac{M_{\rm out}{V'}}{R_{\rm out}})\ , \text{ and}
\end{equation}
\begin{equation}
    \dot{E}_{\rm out} = \frac12\dot{M}_{\rm out}V'^2\ .
\end{equation}
We list the results of the outflow properties in Table~\ref{tab:outflow}. Note that these results largely depend on the assumptions and the procedure of measurements \citep{2023A&A...680A..71H}. The potential outflow components may also be largely submerged in the noise for most of the NIR emission lines in this work.

\begin{table*}[]
\centering
\caption{Properties of the ionized and {warm molecular} gas and outflows {in ID~2}.}
\renewcommand{\arraystretch}{1.5}
\setlength{\tabcolsep}{3mm}{}
\begin{tabular}{ccccccccc}
\hline
\hline
ID  & phase & $\log M_{\rm gas}$ & $f$ & $R_{\rm out}$ & ${V'}$ & $\log M_{\rm out}$ & $\log \dot{M}_{\rm out}$ & $\log \dot{E}_{\rm out}$ \\
&&($ \rm M_\odot$ )&& ($\rm kpc$) &($\rm km/s$) & ($\rm M_\odot$) & ($\rm M_\odot\ yr^{-1}$) & ($\rm erg\ s^{-1}$) \\
(1)&(2)&(3)&(4)&(5)&(6)&(7)&(8)&(9)\\
\hline

2 & ionized & $6.90^{+0.01}_{-0.00}$ & $0.34$ & $0.66^{+0.02}_{-0.02}$ & $727.48^{+38.73}_{-33.14}$ & $6.43^{+0.02}_{-0.02}$ & $0.96^{+0.03}_{-0.03}$ & $42.18^{+0.07}_{-0.07}$ \\
2 & warm molecular & $3.82 ^{+0.03}_{-0.03}$ & $0.20$ & $0.87 ^{+0.20}_{-0.20}$ & $399.08 ^{+45.22}_{-32.50}$ & $3.11 ^{+0.62}_{-0.29}$ & $-2.68 ^{+0.56}_{-0.37}$ & $38.05 ^{+0.55}_{-0.42}$ \\
\hline
    \end{tabular}
    \raggedright
    \tablecomments{(1) Sample ID; (2) the gas phase, either {warm} molecular or ionized gas; (3) the mass of {warm} molecular or ionized gas estimated by the flux of the line; (4) the outflow fraction {$f = \frac{M_{\rm out}}{M_{\rm gas}}$}; (5) the mean value of the radius of the bulk of outflow; (6) the mean value of the velocity of the bulk of outflow; (7) the total mass of the bulk of outflow; (8) the mass outflow rate of the outflow; (9) the outflow {power}. }
    \label{tab:outflow}
\end{table*}

\begin{figure}[ht!]
    \centering
    \includegraphics[scale=0.55]{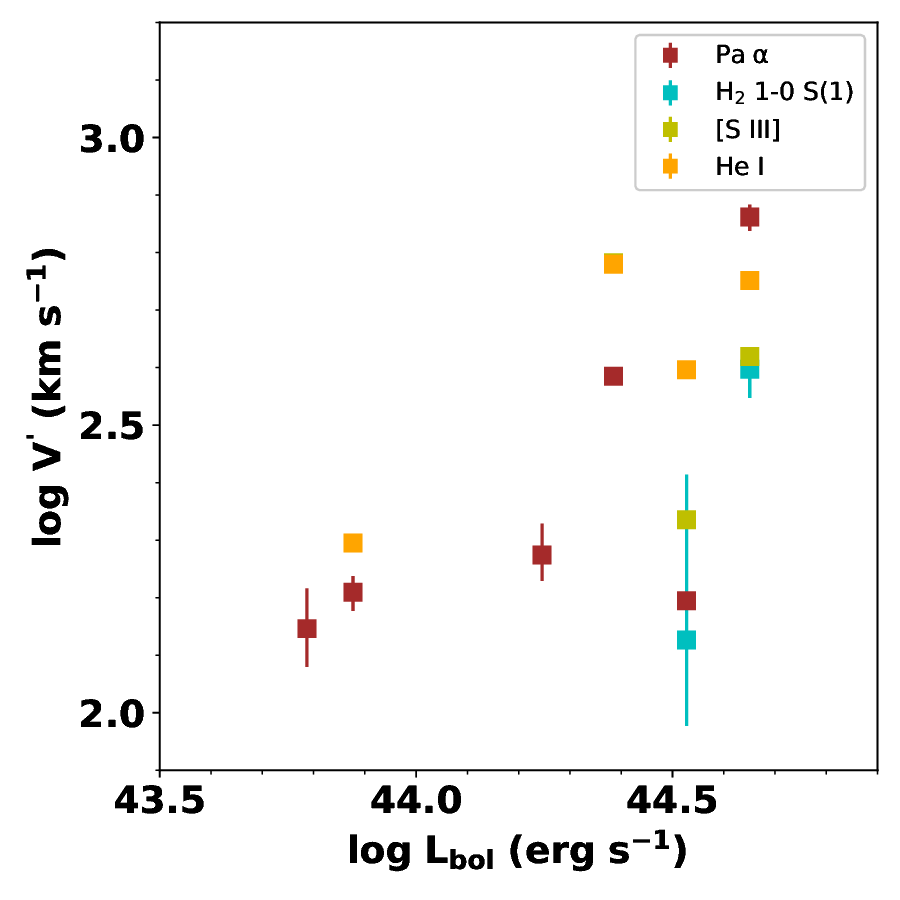}
    \caption{{Maximum outflow} velocity estimated from different emission lines as a function of bolometric luminosity.
    }
    \label{fig:lv}
\end{figure}
\begin{figure}[th!]
    \centering
    \includegraphics[scale=0.32]{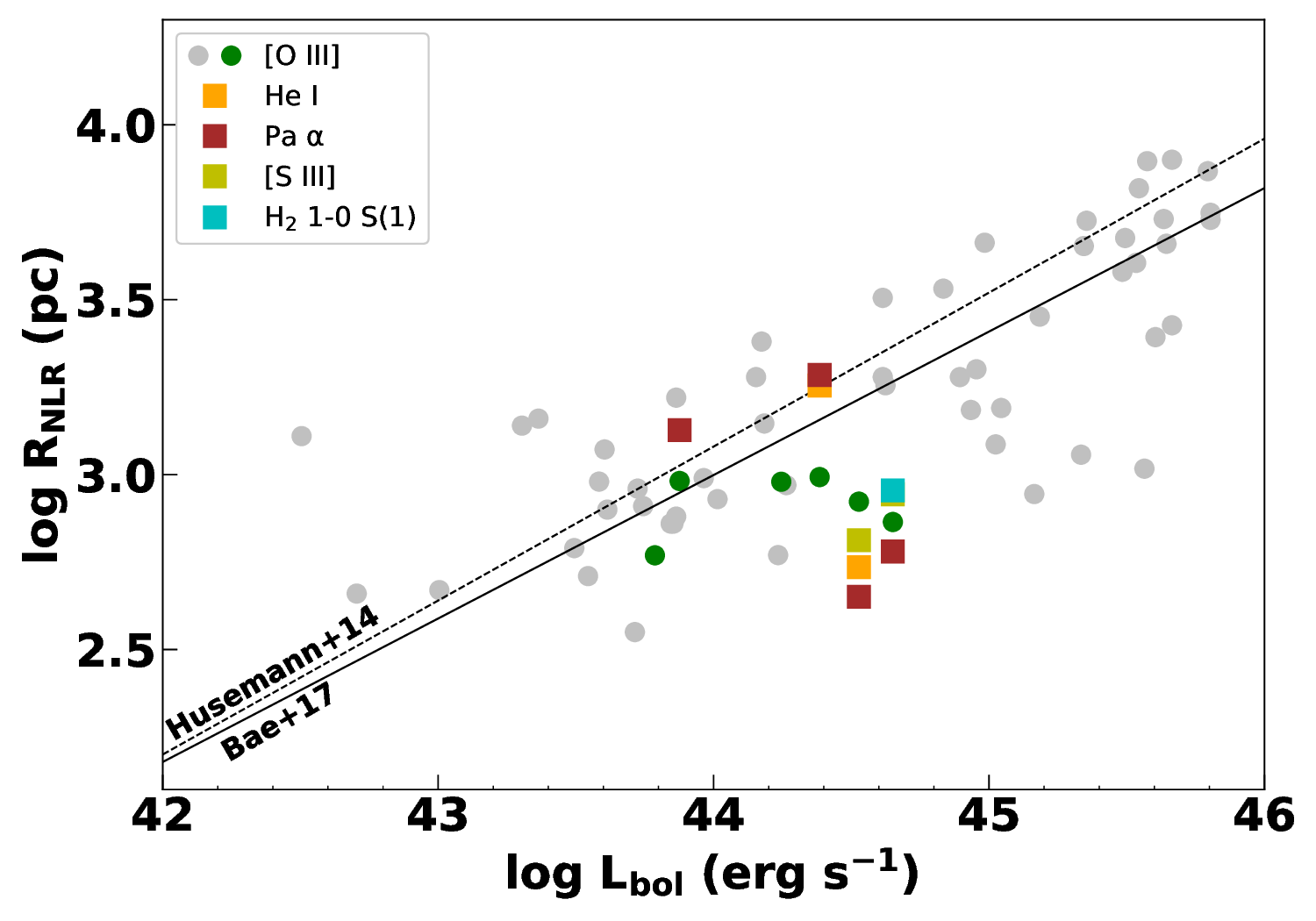}
    \caption{NLR size based on different emission lines as a function of bolometric {luminosity}. The dots show the measured sizes based on optical [O III] of the collected samples from \citet{2013A&A...549A..43H}, \citet{2014MNRAS.443..755H}, \citet{karouzos_unraveling_2016} and \citet{bae_limited_2017}. The colors mark the targets of our sample. Dashed and solid lines present the slope values from the linear regression of \citet{2014MNRAS.443..755H} and \citet{bae_limited_2017}{, respectively}.}
    \label{fig:rl}
\end{figure}
\begin{figure*}[!ht]
    \centering
    \includegraphics[scale=0.55]{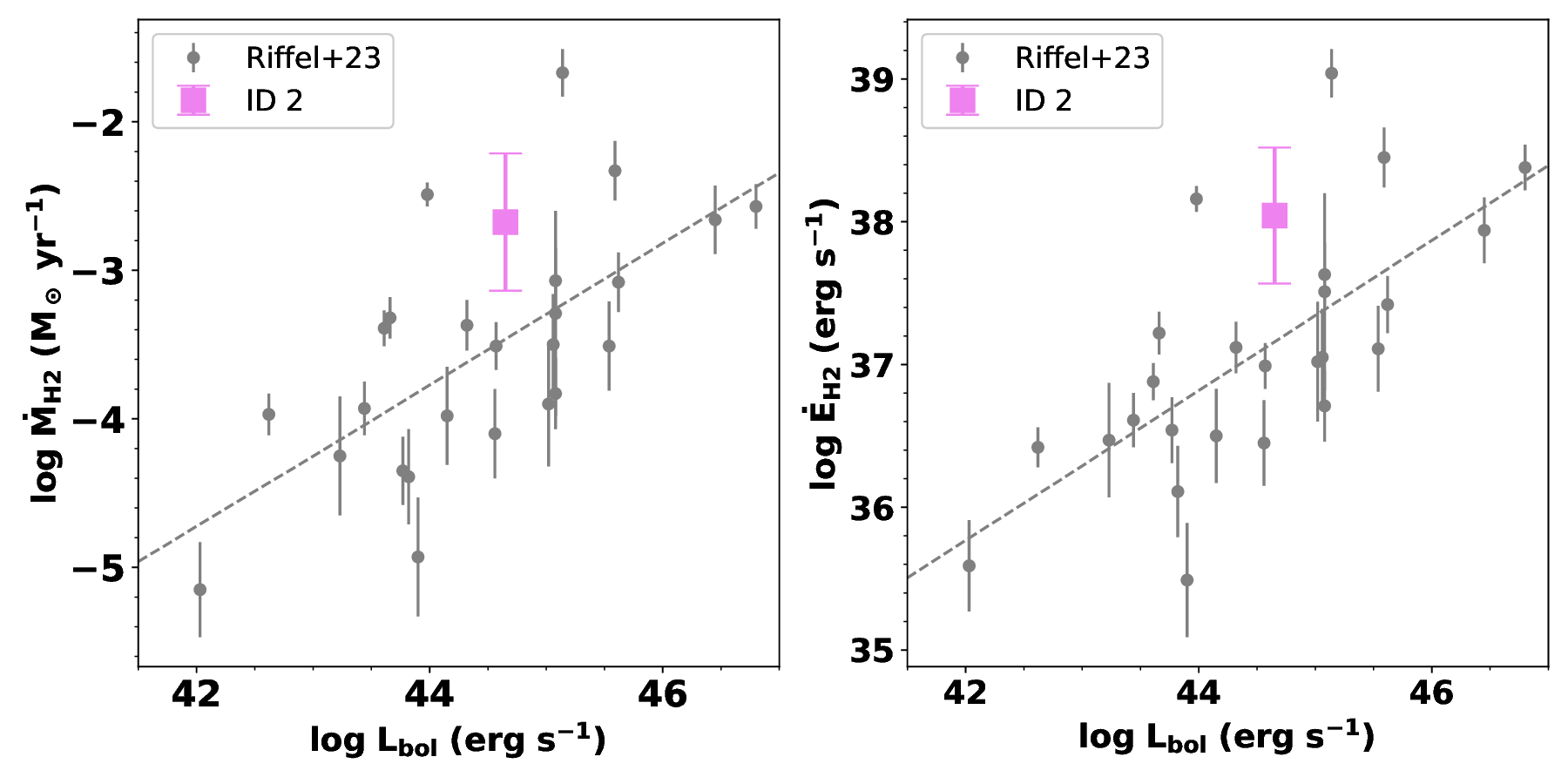}
    \caption{Left panel: plot of mass-outflow rates vs. AGN bolometric luminosity. {Gray points} are outflow properties obtained from \citet{riffel_agnifs_2023}. The dashed line shows a simple linear fit for the {gray points}. The {violet} square notes our results of H$_2$ 1-0 S(1) outflow of ID 2, under the same assumption of spherical geometry. Right panel: same as the left panel, but for the kinetic power of the $\rm H_2$ outflow.}
    \label{fig:me-l}
\end{figure*}
\begin{figure*}[ht!]
    \centering
    \includegraphics[scale=0.4]{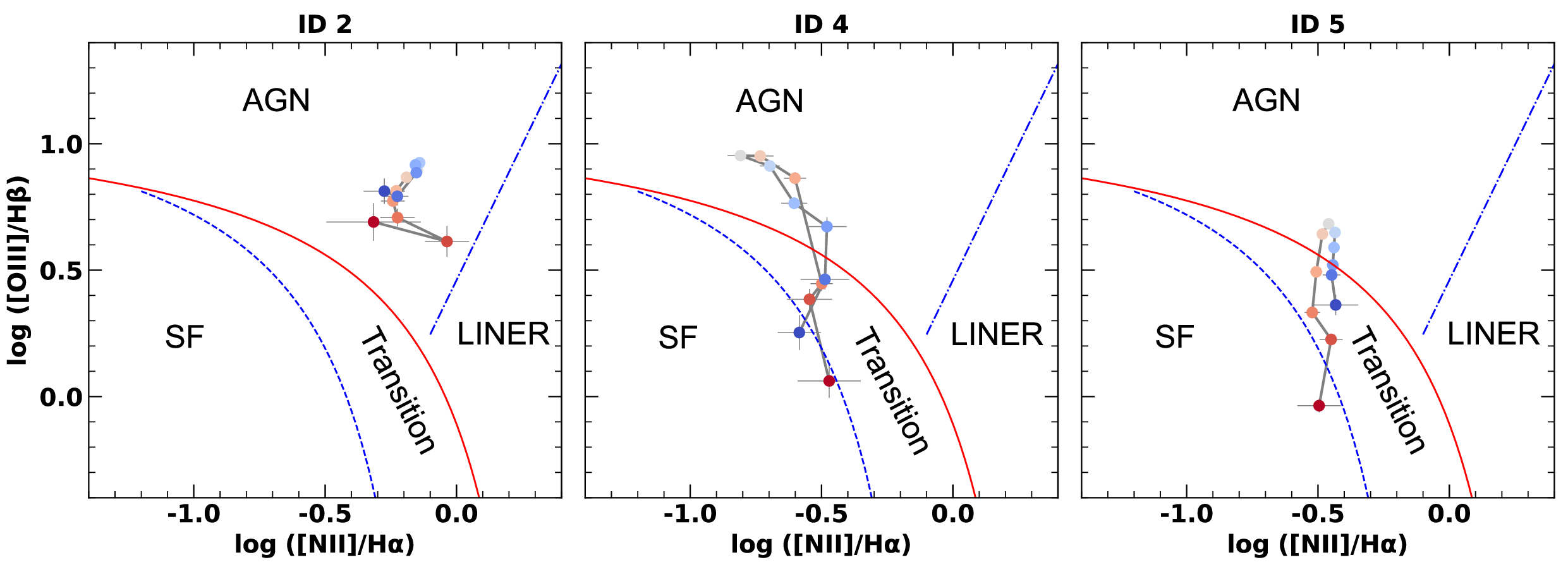}
    \caption{Spatially resolved BPT Diagram for ID 2, ID 4, and ID 5, respectively. {The measurements are derived from the DBSP observations. Lighter colors indicate spectral pixels closer to the centers of the sources. The regions corresponding to AGN, star-forming (SF), transition, and Low-Ionization Nuclear Emission-line Regions (LINERs) are labeled. The spatial step between consecutive points is $0.389 \arcsec$, corresponding to 390, 352, and 471 pc for ID~2, ID~4, and ID~5, respectively.}
}
    \label{fig:bpt}
\end{figure*}
\section{Discussions}\label{sec:dis}

\subsection{AGN Strength and Kinematics}

Despite large scatters, outflows with the highest velocities were found to be driven by the most luminous {AGNs} \citep[e.g.,][]{2017A&A...601A.143F, 2023ApJ...951....7A}. 
{This correlation can be seen in the NIR ionized lines in our sample (as shown in Figure~\ref{fig:lv}), whereas the data for the {warm} molecular line is too limited to discern a trend.
The Spearman correlation coefficient, estimated using \emph{scipy.stats.spearmanr}, is 0.63 (0.54) with an associated $p$-value of 0.02 (0.04) for the ionized lines (both the ionized and {warm} molecular lines).
Note that $V'$ here indicates the outflow velocity only if the outflow exists. If no outflow is present, it merely serves as a parameter to describe the kinematics of the gas.}
{We confirm that if we use only the velocity dispersion $\sigma$ of these lines, which dominates the value of $V'$, the positive correlation persists, with a correlation coefficient of 0.68 (0.57) and a $p$-value of 0.01 (0.03) for the ionized lines (both the ionized and warm molecular lines).} 
However, our selected sample is small and has a narrow range of luminosity. 
The analysis of a larger sample is necessary to confirm this relation.

\subsection{AGN Strength and Narrow-line Region Size}\label{subsec:NRL_size}

Size-luminosity relation has been found in NLR size $R_{\rm NLR}$ and $\rm [O\ III]$ luminosity $L_{[\rm O\ III]}$ \citep{le_ionized-gas_2017}.
We present the NLR sizes of our sample estimated based on spatially resolved emission lines, including both ionized and {warm molecular} gases in Figure~\ref{fig:rl}. 
The archival {samples} are collected from \citet{2013A&A...549A..43H}, \citet{2014MNRAS.443..755H}, \citet{karouzos_unraveling_2016}, \citet{bae_limited_2017}, and \citet{le_ionized-gas_2017}.
For coherence with the comparisons, we take the bolometric luminosity (calculated from extinction-uncorrected [O III] luminosity, $L_{\rm bol} = 3500\times L_{\rm [O\ III]}$; \citealt{2004ApJ...613..109H}) as an indicator of AGN strength.

We find that ionized lines exhibit similar characteristics in this diagram, whether observed in the infrared or optical. 
We note that only two H$_2$ 1-0 S(1) lines have been detected, and thus we cannot draw a conclusive trend for {warm} molecular lines.
We cannot see a clear distinction between [O III] and other lines, as we expect [S III], He I, $\rm H_2$ and $\rm Pa\alpha$ to be more extended. This may be attributed to the noise that leads to an underestimation of the sizes.
{It is important to note that the derived radii of the sizes may be underestimated, as the low-surface-brightness component submerged by the noise was not considered in our adopted methodology \citep{2022A&A...665A..55S, 2024A&A...681A..63S}.
Therefore, our reported $R_{\rm NLR}$, as well as $R_{\rm out}$, should be considered as conservative estimates or lower limits of the true physical extent of the ionized gas outflows.
}



\subsection{AGN Strength and Outflow Properties}
{To characterize outflow sizes,} \citet{karouzos_unraveling_2016} estimated the radius encompassing half of the total broad $\rm [O\ III]$ emission flux as the effective radius $r_{\rm eff}$. In the case of ID 2 in our sample, the radius of the bulk of the ionized outflow is $0.66 ^{+0.02}_{-0.02}\ \rm kpc$, which is similar to the effective radius of $0.60\ \rm kpc$ estimated from [O III] by \citet{karouzos_unraveling_2016}.

Generally, {warm molecular} outflows are anticipated to have a larger radius compared to the ionized {outflows} \citep{karouzos_unraveling_2016}. This is because the {warm} molecular component of the outflow originates from the cooling of gas confined within the outflow. As shown in Table~\ref{tab:outflow},  {the size of the warm molecular outflow ($0.87^{+0.20}_{-0.20}~\rm kpc$) is larger than that of the ionized gas, although the difference is only at the level of one to two standard deviations.
This may be attributed to the limited S/N and the measurement neglecting the low-surface-brightness component, which can lead to non-detection of the weaker and colder molecular gas over a broader region, as discussed in Section~\ref{subsec:NRL_size}.}

Using an X-ray-selected sample, \citet{riffel_agnifs_2023} found that mass outflow rates and {warm molecular} outflow power are positively correlated with the AGN bolometric luminosity. 
As shown in Figure~\ref{fig:me-l}, we compare our only detected {warm molecular} outflow, tracing by H$_2$ 1-0 S(1) of ID 2, with values from the literature \citep{riffel_agnifs_2023}. 
{We employ linear regression to illustrate the positive correlation among these values, and our results are generally consistent with the overall trend, albeit being slightly higher. 
It should be noted that the outflow energetics derived in this work are influenced by the analysis methods. First, as discussed in Section~\ref{subsec:NRL_size}, $R_{\rm out}$ is likely underestimated, which leads to an overestimation of $\dot{M}_{\rm out}$ and $\dot{E}_{\rm out}$. Second, as shown by \citet{2023A&A...680A..71H}, using maximum outflow velocities derived from parametric fits tends to produce higher estimates compared to other methods. 
In addition, the assumption of a constant density also affects the estimated properties of ionized outflows, since adopting lower densities would lead to an overestimation of the outflow mass and energy.
These may contribute to the deviation from the result and the overall trend.
}

\subsection{Spatially Resolved BPT Diagram}
The \citeauthor*{1981PASP...93....5B} (BPT) diagram can be used to determine whether a spectrum is contributed from AGN or stellar components. Due to the difference in angular distance of 1 pixel between the red and blue cameras, we utilized the $\rm [N\ II]/H\alpha$ line ratio fitted from the red 2D spectra using B-spline interpolation to estimate the $\rm [N\ II]/H\alpha$ line ratio at the corresponding position in the blue 2D spectra of DBSP. 
{Therefore, the spatial step between consecutive points is $0.389 \arcsec$, corresponding to 390, 352, and 471 pc for ID~2, ID~4, and ID~5, respectively.}

Figure~\ref{fig:bpt} illustrates the BPT diagrams for the three sources ID~2, ID~4, and ID~5, {where lighter colors indicate spectral pixels closer to the center of the sources}. 
We note that for ID~4 and ID~5, the spectra corresponding to several pixels in the center fall within the AGN region on the BPT diagram, and as we move farther away from the center, they become closer to the region associated with stars. 
{The spectra of ID 2 remain located in the AGN region at larger distances from the center, indicating that AGN photoionization or shock excitation continues to dominate the emission even in the outer regions, consistent with stronger AGN activity.}
This indicates that our two-dimensional spectrum possesses a specific spatial resolution capability and offers a method for determining the regions affected by AGN activity.

\subsection{Origin of {Warm Molecular} Gas Emission}

The $\rm H_2/Br\gamma$ line ratio serves as a useful tool for investigating the origin of $\rm H_2$ emission.
{\citet{riffel_agnifs_2021} summarized that the $\mathrm{H_2}$ emission can arise from three main excitation mechanisms: fluorescence ($\mathrm{H_2/Br\gamma < 0.4}$), X-ray heating and shock-related excitation ($0.4 < \mathrm{H_2/Br\gamma} < 6$), and the shock-dominated regime ($\mathrm{H_2/Br\gamma > 6}$).}
We have detected the $\rm H_2$ and $\rm Br\gamma$ emission lines in ID 2, with an integrated $\rm H_2/Br\gamma$ line ratio of approximately $1$. This finding suggests that the $\rm H_2$ emission is primarily driven by thermal processes in ID 2, including X-ray heating and shocks, possibly attributed to AGN activities.

\subsection{Possibility of Detecting Both Ionized and {Warm Molecular} Outflows}

{
Recent studies highlight the complex, multi-phase nature of AGN feedback.
While ionized and cold molecular outflows are frequently detected in such systems \citep[e.g.,][]{rupke_breaking_2013,2022A&A...658A.155R, 2024A&A...681A..63S}, the warm molecular counterpart is not always observed \citep[e.g.,][]{ramos_almeida_infrared_2017}.}
And \citet{riffel_agnifs_2023} found that the ionized outflow was found in 96\% of their sample, while it was 76\% for the {warm molecular} outflow. 
As in our sample, the occurrence of {warm molecular} outflows is notably lower at 16\% (1 out of 6 objects having {warm} molecular outflows). We hypothesize that some potential outflow components might be submerged by noise due to the limited S/N, even for ID 2.
Broad components in [O III] emission lines (see Figure~\ref{fig:vs}) are evident. However, a similar profile is solely detected in $\rm Pa\alpha$. Given the considerably higher S/N of [O III] emission compared to NIR lines, it is plausible that some outflow signatures are missed in this analysis. 
To address this problem, further high-quality and deep spectroscopic observations are needed.
{While we focus here on the ionized and warm molecular phase, a complete picture requires comparison with other gas phases \citep[e.g.,][]{2022A&A...665A..55S}.
Investigating the cold molecular gas phase through radio observations is also valuable for obtaining a more comprehensive understanding of AGN feedback.
In addition, }extensive work on type~1 AGNs is carried out in \citet[][i.e., \citetalias{2026arXiv260110372Q}]{2026arXiv260110372Q}.

\section{Summary}\label{sec:conc}

As a complement to \citetalias{le_active_2024}, we present a study focusing on comparing ionized and {warm molecular} gas outflows in six {type-2} AGNs with redshifts below 0.1, demonstrating extraordinary levels of ionized gas outflows. The study was conducted employing the TPSP and DBSP instruments on the P200 telescope and we supplement our data with archival SDSS data. We summarize the primary findings and crucial conclusions as follows.

\begin{enumerate}

\item The {warm molecular} gas traced by H$_2$ 1-0 S(1) line is detected in two of six AGNs in our sample, with one (ID 2) showing {significant outflow signatures}. The non-detection of the outflow components in the remaining spectra may be due to the limited S/N. However, at least not all objects with strong ionized outflows have strong {warm} molecular outflows.

\item We estimate the radii of ionized and {warm molecular} outflow sizes for ID 2 to be {$0.66 ^{+0.02}_{-0.02}\ \rm kpc$ and $0.87 ^{+0.20}_{-0.20} \rm kpc$}, respectively. The profile of $\rm H_2$ is skewed to the left, indicating {that} the intensity distribution of the {warm molecular} outflow is different from the ionized outflow in ID 2.

\item A positive correlation between kinematics and luminosity is shown in both ionized and {warm} molecular lines, suggesting that more luminous AGNs, which reflect higher levels of AGN activity, tend to have a greater impact on the gas, probably {driving} the outflows.

\item We estimate the outflow rates of mass and energy of the {warm molecular} outflow of ID 2 and compare them with the X-ray-selected sample {presented by \citet{riffel_agnifs_2023}. We find that ID~2 follows the expected positive correlation with the bolometric luminosity established by the comparison sample.}

\item The spatially resolved BPT diagram of the three sources observed by DBSP suggests regions influenced by the AGN activities, providing a different way to measure the AGN strength, probably with a better resolution.

\item We find {that} the integrated $\rm H_2/Br\gamma$ line ratio {is} approximately 1. This finding suggests that the $\rm H_2$ emission is primarily driven by thermal processes in ID 2, including X-ray heating and shocks, possibly attributed to AGN activities.


\end{enumerate}

\begin{acknowledgments}

This work has been supported by the National Natural Science Foundation of China (NSFC-12473014, NSFC-12025303, NSFC-12203047), National Key R\&D Program of China No. 2022YFF0503401, {2023YFA1608100}. 
This research uses data obtained through the Telescope Access Program (TAP), which has been funded by the TAP member institutes (Proposal ID: CTAP2022-A0037). We thank the DBSP observation{al} data shared by Luming Sun and Yibo Wang.
\end{acknowledgments}


\bibliographystyle{aasjournal}
\bibliography{tspec}

@article{le_fluorescent_2017,
	title = {Fluorescent {H}\_2 {Emission} {Lines} from the {Reflection} {Nebula} {NGC} 7023 {Observed} with {IGRINS}},
	volume = {841},
	issn = {1538-4357},
	url = {http://arxiv.org/abs/1609.01818},
	doi = {10.3847/1538-4357/aa6bf7},
	number = {1},
	urldate = {2022-02-21},
	journal = {The Astrophysical Journal},
	author = {Le, Huynh Anh N. and Pak, Soojong and Kaplan, Kyle F. and Mace, Gregory N. and Lee, Sungho and Pavel, Michael D. and Jeong, Ueejeong and Oh, Heeyoung and Lee, Hye-In and Chun, Moo-Young and Yuk, In-Soo and Pyo, Tae-Soo and Hwang, Narae and Kim, Kang-Min and Park, Chan and Oh, Jae Sok and Yu, Young S. and Park, Byeong-Gon and Minh, Young Chol and Jaffe, Daniel T.},
	month = may,
	year = {2017},
	note = {arXiv: 1609.01818},
	pages = {13},
}

@article{bae_census_2014,
	title = {A {CENSUS} {OF} {GAS} {OUTFLOWS} {IN} {TYPE} 2 {ACTIVE} {GALACTIC} {NUCLEI}},
	language = {en},
	journal = {The Astrophysical Journal},
	author = {Bae, Hyun-Jin and Woo, Jong-Hak},
	year = {2014},
	pages = {11},
}

@article{riffel_0824_2006,
	title = {A 0.8–2.4 \textit{μ} m spectral atlas of active galactic nuclei},
	volume = {457},
	issn = {0004-6361, 1432-0746},
	url = {http://www.aanda.org/10.1051/0004-6361:20065291},
	doi = {10.1051/0004-6361:20065291},
	language = {en},
	number = {1},
	urldate = {2022-02-26},
	journal = {Astronomy \& Astrophysics},
	author = {Riffel, R. and Rodríguez-Ardila, A. and Pastoriza, M. G.},
	month = oct,
	year = {2006},
	pages = {61--70},
}

@article{ramosalmeida_near-infrared_2019,
	title = {A near-infrared study of the multiphase outflow in the type-2 quasar {J1509}+0434},
	volume = {487},
	issn = {1745-3925},
	url = {https://doi.org/10.1093/mnrasl/slz072},
	doi = {10.1093/mnrasl/slz072},
	number = {1},
	urldate = {2022-03-28},
	journal = {Monthly Notices of the Royal Astronomical Society: Letters},
	author = {Ramos Almeida, C and Acosta-Pulido, J A and Tadhunter, C N and González-Fernández, C and Cicone, C and Fernández-Torreiro, M},
	month = jul,
	year = {2019},
	pages = {L18--L23},
}

@article{bianchin_gemini_2022,
	title = {Gemini {NIFS} survey of feeding and feedback in nearby active galaxies - {V}. {Molecular} and ionized gas kinematics},
	volume = {510},
	issn = {0035-8711},
	url = {https://ui.adsabs.harvard.edu/abs/2022MNRAS.510..639B/abstract},
	doi = {10.1093/mnras/stab3468},
	language = {en},
	number = {1},
	urldate = {2022-06-23},
	journal = {Monthly Notices of the Royal Astronomical Society},
	author = {Bianchin, M. and Riffel, R. A. and Storchi-Bergmann, T. and Riffel, R. and Ruschel-Dutra, D. and Harrison, C. M. and Dahmer-Hahn, L. G. and Mainieri, V. and Schönell, A. J. and Dametto, N. Z.},
	month = feb,
	year = {2022},
	pages = {639--657},
}

@article{karouzos_unraveling_2016,
	title = {{UNRAVELING} {THE} {COMPLEX} {STRUCTURE} {OF} {AGN}-{DRIVEN} {OUTFLOWS}. {I}. {KINEMATICS} {AND} {SIZES}},
	volume = {819},
	issn = {1538-4357},
	url = {https://iopscience.iop.org/article/10.3847/0004-637X/819/2/148},
	doi = {10.3847/0004-637X/819/2/148},
	language = {en},
	number = {2},
	urldate = {2022-07-17},
	journal = {The Astrophysical Journal},
	author = {Karouzos, Marios and Woo, Jong-Hak and Bae, Hyun-Jin},
	month = mar,
	year = {2016},
	pages = {148},
}

@article{riffel_gemini_2021,
	title = {Gemini {NIFS} survey of feeding and feedback in nearby active galaxies – {IV}. {Excitation}},
	volume = {503},
	issn = {0035-8711},
	url = {https://doi.org/10.1093/mnras/stab788},
	doi = {10.1093/mnras/stab788},
	number = {4},
	urldate = {2022-07-17},
	journal = {Monthly Notices of the Royal Astronomical Society},
	author = {Riffel, Rogemar A and Bianchin, Marina and Riffel, Rogério and Storchi-Bergmann, Thaisa and Schönell, Astor J and Dahmer-Hahn, Luis Gabriel and Dametto, Natacha Z and Diniz, Marlon R},
	month = jun,
	year = {2021},
	pages = {5161--5178},
}

@article{scoville_velocity_1982,
	title = {Velocity, reddening, and temperature structure of the {H2} emission in {Orion}},
	volume = {253},
	issn = {0004-637X},
	url = {https://ui.adsabs.harvard.edu/abs/1982ApJ...253..136S},
	doi = {10.1086/159618},
	urldate = {2022-08-05},
	journal = {The Astrophysical Journal},
	author = {Scoville, N. Z. and Hall, D. N. B. and Ridgway, S. T. and Kleinmann, S. G.},
	month = feb,
	year = {1982},
	note = {ADS Bibcode: 1982ApJ...253..136S},
	pages = {136--148},
}

@article{riffel_feeding_2014,
	title = {Feeding versus feedback in {NGC} 1068 probed with {Gemini} {NIFS} – {I}. {Excitation}},
	volume = {442},
	issn = {0035-8711},
	url = {https://doi.org/10.1093/mnras/stu843},
	doi = {10.1093/mnras/stu843},
	number = {1},
	urldate = {2022-09-28},
	journal = {Monthly Notices of the Royal Astronomical Society},
	author = {Riffel, Rogemar A. and Vale, Tiberio B. and Storchi-Bergmann, Thaisa and McGregor, Peter J.},
	month = jul,
	year = {2014},
	pages = {656--669},
}

@article{rose_quantifying_2018,
	title = {Quantifying the {AGN}-driven outflows in {ULIRGs} ({QUADROS}) – {I}: {VLT}/{Xshooter} observations of nine nearby objects},
	volume = {474},
	issn = {0035-8711, 1365-2966},
	shorttitle = {Quantifying the {AGN}-driven outflows in {ULIRGs} ({QUADROS}) – {I}},
	url = {http://academic.oup.com/mnras/article/474/1/128/4349765},
	doi = {10.1093/mnras/stx2590},
	language = {en},
	number = {1},
	urldate = {2022-12-02},
	journal = {Monthly Notices of the Royal Astronomical Society},
	author = {Rose, Marvin and Tadhunter, Clive and Ramos Almeida, Cristina and Rodríguez Zaurín, Javier and Santoro, Francesco and Spence, Robert},
	month = feb,
	year = {2018},
	pages = {128--156},
}

@article{kormendy_coevolution_2013,
	title = {Coevolution ({Or} {Not}) of {Supermassive} {Black} {Holes} and {Host} {Galaxies}},
	volume = {51},
	issn = {0066-4146},
	url = {https://ui.adsabs.harvard.edu/abs/2013ARA&A..51..511K},
	doi = {10.1146/annurev-astro-082708-101811},
	urldate = {2022-12-08},
	journal = {Annual Review of Astronomy and Astrophysics},
	author = {Kormendy, John and Ho, Luis C.},
	month = aug,
	year = {2013},
	note = {ADS Bibcode: 2013ARA\&A..51..511K},
	pages = {511--653},
}

@article{woo_prevalence_2016,
	title = {The {Prevalence} of {Gas} {Outflows} in {Type} 2 {AGNs}},
	volume = {817},
	issn = {0004-637X},
	url = {https://ui.adsabs.harvard.edu/abs/2016ApJ...817..108W},
	doi = {10.3847/0004-637X/817/2/108},
	urldate = {2022-12-13},
	journal = {The Astrophysical Journal},
	author = {Woo, Jong-Hak and Bae, Hyun-Jin and Son, Donghoon and Karouzos, Marios},
	month = feb,
	year = {2016},
	note = {ADS Bibcode: 2016ApJ...817..108W},
	pages = {108},
}

@article{harrison_agn_2018,
	title = {{AGN} outflows and feedback twenty years on},
	volume = {2},
	issn = {2397-3366},
	url = {http://arxiv.org/abs/1802.10306},
	doi = {10.1038/s41550-018-0403-6},
	number = {3},
	urldate = {2023-01-13},
	journal = {Nature Astronomy},
	author = {Harrison, C. M. and Costa, T. and Tadhunter, C. N. and Flütsch, A. and Kakkad, D. and Perna, M. and Vietri, G.},
	month = feb,
	year = {2018},
	note = {arXiv:1802.10306 [astro-ph]},
	pages = {198--205},
}

@article{ramos_almeida_infrared_2017,
	title = {An infrared view of {AGN} feedback in a type-2 quasar: the case of the {Teacup} galaxy},
	volume = {470},
	issn = {0035-8711, 1365-2966},
	shorttitle = {An infrared view of {AGN} feedback in a type-2 quasar},
	url = {https://academic.oup.com/mnras/article/470/1/964/3852307},
	doi = {10.1093/mnras/stx1287},
	language = {en},
	number = {1},
	urldate = {2023-03-23},
	journal = {Monthly Notices of the Royal Astronomical Society},
	author = {Ramos Almeida, C. and Piqueras López, J. and Villar-Martín, M. and Bessiere, P. S.},
	month = sep,
	year = {2017},
	pages = {964--976},
}

@article{riffel_agnifs_2023,
	title = {The {AGNIFS} survey: spatially resolved observations of hot molecular and ionised outflows in nearby active galaxies},
	volume = {521},
	issn = {0035-8711, 1365-2966},
	shorttitle = {The {AGNIFS} survey},
	url = {http://arxiv.org/abs/2302.11324},
	doi = {10.1093/mnras/stad599},
	language = {en},
	number = {2},
	urldate = {2023-04-06},
	journal = {Monthly Notices of the Royal Astronomical Society},
	author = {Riffel, R. A. and Storchi-Bergmann, T. and Riffel, R. and Bianchin, M. and Zakamska, N. L. and Ruschel-Dutra, D. and Bentz, M. C. and Burtscher, L. and Crenshaw, D. M. and Dahmer-Hahn, L. G. and Dametto, N. Z. and Davies, R. I. and Diniz, M. R. and Fischer, T. C. and Harrison, C. M. and Mainieri, V. and Revalski, M. and Rodriguez-Ardila, A. and Rosario, D. J. and Schonell, A. J.},
	month = mar,
	year = {2023},
	note = {arXiv:2302.11324 [astro-ph]},
	pages = {1832--1848},
}

@article{riffel_agnifs_2021,
	title = {The {AGNIFS} survey: distribution and excitation of the hot molecular and ionised gas in the inner kpc of nearby {AGN} hosts},
	volume = {504},
	issn = {0035-8711, 1365-2966},
	shorttitle = {The {AGNIFS} survey},
	url = {http://arxiv.org/abs/2104.03105},
	doi = {10.1093/mnras/stab998},
	number = {3},
	urldate = {2023-06-05},
	journal = {Monthly Notices of the Royal Astronomical Society},
	author = {Riffel, R. A. and Storchi-Bergmann, T. and Riffel, R. and Bianchin, M. and Zakamska, N. L. and Ruschel-Dutra, D. and Schonell, A. J. and Rosario, D. J. and Rodriguez-Ardila, A. and Fischer, T. C. and Davies, R. I. and Dametto, N. Z. and Dahmer-Hahn, L. G. and Crenshaw, D. M. and Burtscher, L. and Bentz, M. C.},
	month = may,
	year = {2021},
	note = {arXiv:2104.03105 [astro-ph]},
	pages = {3265--3283},
}

@article{zubovas_galaxy-wide_2014,
	title = {Galaxy-wide outflows: cold gas and star formation at high speeds},
	volume = {439},
	issn = {0035-8711},
	shorttitle = {Galaxy-wide outflows},
	url = {https://ui.adsabs.harvard.edu/abs/2014MNRAS.439..400Z},
	doi = {10.1093/mnras/stt2472},
	urldate = {2023-06-13},
	journal = {Monthly Notices of the Royal Astronomical Society},
	author = {Zubovas, Kastytis and King, Andrew R.},
	month = mar,
	year = {2014},
	note = {ADS Bibcode: 2014MNRAS.439..400Z},
	pages = {400--406},
}

@article{bae_prevalence_2016,
	title = {The {Prevalence} of {Gas} {Outflows} in {Type} 2 {AGNs}. {II}. {3D} {Biconical} {Outflow} {Models}},
	volume = {828},
	issn = {1538-4357},
	url = {http://arxiv.org/abs/1606.05348},
	doi = {10.3847/0004-637X/828/2/97},
	number = {2},
	urldate = {2023-06-13},
	journal = {The Astrophysical Journal},
	author = {Bae, Hyun-Jin and Woo, Jong-Hak},
	month = sep,
	year = {2016},
	note = {arXiv:1606.05348 [astro-ph]},
	pages = {97},
}

@article{riffel_molecular_2013,
	title = {Molecular hydrogen and [{Fe} {II}] in active galactic nuclei - {III}. {Low}-ionization nuclear emission-line region and star-forming galaxies},
	volume = {430},
	issn = {0035-8711},
	url = {https://ui.adsabs.harvard.edu/abs/2013MNRAS.430.2002R},
	doi = {10.1093/mnras/stt026},
	urldate = {2023-06-13},
	journal = {Monthly Notices of the Royal Astronomical Society},
	author = {Riffel, R. and Rodríguez-Ardila, A. and Aleman, I. and Brotherton, M. S. and Pastoriza, M. G. and Bonatto, C. and Dors, O. L.},
	month = apr,
	year = {2013},
	note = {ADS Bibcode: 2013MNRAS.430.2002R},
	pages = {2002--2017},
}

@article{riffel_active_2020,
	title = {Active galactic nuclei winds as the origin of the {H2} emission excess in nearby galaxies},
	volume = {491},
	issn = {0035-8711},
	url = {https://ui.adsabs.harvard.edu/abs/2020MNRAS.491.1518R},
	doi = {10.1093/mnras/stz3137},
	urldate = {2023-06-13},
	journal = {Monthly Notices of the Royal Astronomical Society},
	author = {Riffel, Rogemar A. and Zakamska, Nadia L. and Riffel, Rogério},
	month = jan,
	year = {2020},
	note = {ADS Bibcode: 2020MNRAS.491.1518R},
	pages = {1518--1529},
}

@article{storchi-bergmann_feeding_2009,
	title = {Feeding versus feedback in {NGC4151} probed with {Gemini} {NIFS} - {I}. {Excitation}},
	volume = {394},
	issn = {0035-8711},
	url = {https://ui.adsabs.harvard.edu/abs/2009MNRAS.394.1148S},
	doi = {10.1111/j.1365-2966.2009.14388.x},
	urldate = {2023-06-13},
	journal = {Monthly Notices of the Royal Astronomical Society},
	author = {Storchi-Bergmann, T. and McGregor, P. J. and Riffel, Rogemar A. and Simões Lopes, R. and Beck, T. and Dopita, M.},
	month = apr,
	year = {2009},
	note = {ADS Bibcode: 2009MNRAS.394.1148S},
	pages = {1148--1166},
}

@article{rupke_breaking_2013,
	title = {Breaking the {Obscuring} {Screen}: {A} {Resolved} {Molecular} {Outflow} in a {Buried} {QSO}},
	volume = {775},
	issn = {0004-637X},
	shorttitle = {Breaking the {Obscuring} {Screen}},
	url = {https://ui.adsabs.harvard.edu/abs/2013ApJ...775L..15R},
	doi = {10.1088/2041-8205/775/1/L15},
	urldate = {2023-06-13},
	journal = {The Astrophysical Journal},
	author = {Rupke, David S. N. and Veilleux, Sylvain},
	month = sep,
	year = {2013},
	note = {ADS Bibcode: 2013ApJ...775L..15R},
	pages = {L15},
}

@article{xue_chandra_2017,
	title = {The {Chandra} deep fields: {Lifting} the veil on distant active galactic nuclei and {X}-ray emitting galaxies},
	volume = {79},
	issn = {13876473},
	shorttitle = {The {Chandra} deep fields},
	url = {https://linkinghub.elsevier.com/retrieve/pii/S138764731730026X},
	doi = {10.1016/j.newar.2017.09.002},
	language = {en},
	urldate = {2023-06-13},
	journal = {New Astronomy Reviews},
	author = {Xue, Y.Q.},
	month = nov,
	year = {2017},
	pages = {59--84},
}

@article{u_inner_2013,
	title = {The {Inner} {Kiloparsec} of {Mrk} 273 with {Keck} {Adaptive} {Optics}},
	volume = {775},
	issn = {0004-637X},
	url = {https://ui.adsabs.harvard.edu/abs/2013ApJ...775..115U},
	doi = {10.1088/0004-637X/775/2/115},
	urldate = {2023-06-13},
	journal = {The Astrophysical Journal},
	author = {U, Vivian and Medling, Anne and Sanders, David and Max, Claire and Armus, Lee and Iwasawa, Kazushi and Evans, Aaron and Kewley, Lisa and Fazio, Giovanni},
	month = oct,
	year = {2013},
	note = {ADS Bibcode: 2013ApJ...775..115U},
	pages = {115},
}

@article{izotov_near-infrared_2016,
	title = {Near-infrared spectroscopy of a large sample of low-metallicity blue compact dwarf galaxies},
	volume = {457},
	issn = {0035-8711},
	url = {https://ui.adsabs.harvard.edu/abs/2016MNRAS.457...64I},
	doi = {10.1093/mnras/stv2957},
	urldate = {2023-06-14},
	journal = {Monthly Notices of the Royal Astronomical Society},
	author = {Izotov, Y. I. and Thuan, T. X.},
	month = mar,
	year = {2016},
	note = {ADS Bibcode: 2016MNRAS.457...64I},
	pages = {64--73},
}

@article{prochaska_pypeit_2020,
	title = {{PypeIt}: {The} {Python} {Spectroscopic} {Data} {Reduction} {Pipeline}},
	volume = {5},
	shorttitle = {{PypeIt}},
	url = {https://ui.adsabs.harvard.edu/abs/2020JOSS....5.2308P},
	doi = {10.21105/joss.02308},
	urldate = {2023-06-14},
	journal = {The Journal of Open Source Software},
	author = {Prochaska, J. and Hennawi, Joseph and Westfall, Kyle and Cooke, Ryan and Wang, Feige and Hsyu, Tiffany and Davies, Frederick and Farina, Emanuele and Pelliccia, Debora},
	month = dec,
	year = {2020},
	note = {ADS Bibcode: 2020JOSS....5.2308P},
	pages = {2308},
}

@book{rice_mathematical_2007,
	address = {Belmont, CA},
	edition = {3rd ed},
	series = {Duxbury advanced series},
	title = {Mathematical statistics and data analysis},
	isbn = {978-0-534-39942-9},
	language = {en},
	publisher = {Thomson/Brooks/Cole},
	author = {Rice, John A.},
	year = {2007},
}

@article{ruschel-dutra_agnifs_2021,
	title = {{AGNIFS} survey of local {AGN}: {GMOS}-{IFU} data and outflows in 30 sources},
	volume = {507},
	issn = {0035-8711},
	shorttitle = {{AGNIFS} survey of local {AGN}},
	url = {https://ui.adsabs.harvard.edu/abs/2021MNRAS.507...74R},
	doi = {10.1093/mnras/stab2058},
	urldate = {2023-06-17},
	journal = {Monthly Notices of the Royal Astronomical Society},
	author = {Ruschel-Dutra, D. and Storchi-Bergmann, T. and Schnorr-Müller, A. and Riffel, R. A. and Dall'Agnol de Oliveira, B. and Lena, D. and Robinson, A. and Nagar, N. and Elvis, M.},
	month = oct,
	year = {2021},
	note = {ADS Bibcode: 2021MNRAS.507...74R},
	pages = {74--89},
}

@article{le_ionized-gas_2017,
	title = {Ionized-gas {Kinematics} {Along} the {Large}-scale {Radio} {Jets} in {Type}-2 {AGNs}},
	volume = {851},
	issn = {0004-637X},
	url = {https://ui.adsabs.harvard.edu/abs/2017ApJ...851....8L},
	doi = {10.3847/1538-4357/aa9656},
	urldate = {2023-06-24},
	journal = {The Astrophysical Journal},
	author = {Le, Huynh Anh N. and Woo, Jong-Hak and Son, Donghoon and Karouzos, Marios and Chung, Aeree and Jung, Taehyun and Tremou, Evangelia and Hwang, Narae and Park, Byeong-Gon},
	month = dec,
	year = {2017},
	note = {ADS Bibcode: 2017ApJ...851....8L},
	pages = {8},
}

@article{bae_limited_2017,
	title = {The {Limited} {Impact} of {Outflows}: {Integral}-field {Spectroscopy} of 20 {Local} {AGNs}},
	volume = {837},
	issn = {0004-637X},
	shorttitle = {The {Limited} {Impact} of {Outflows}},
	url = {https://ui.adsabs.harvard.edu/abs/2017ApJ...837...91B},
	doi = {10.3847/1538-4357/aa5f5c},
	urldate = {2023-06-25},
	journal = {The Astrophysical Journal},
	author = {Bae, Hyun-Jin and Woo, Jong-Hak and Karouzos, Marios and Gallo, Elena and Flohic, Helene and Shen, Yue and Yoon, Suk-Jin},
	month = mar,
	year = {2017},
	note = {ADS Bibcode: 2017ApJ...837...91B},
	pages = {91},
}

@article{kaastra_optimal_2016,
	title = {Optimal binning of {X}-ray spectra and response matrix design},
	volume = {587},
	issn = {0004-6361, 1432-0746},
	url = {http://www.aanda.org/10.1051/0004-6361/201527395},
	doi = {10.1051/0004-6361/201527395},
	language = {en},
	urldate = {2023-07-31},
	journal = {Astronomy \& Astrophysics},
	author = {Kaastra, J. S. and Bleeker, J. A. M.},
	month = mar,
	year = {2016},
	pages = {A151},
}

@misc{cappellari_full_2022,
	title = {Full spectrum fitting with photometry in ppxf: non-parametric star formation history, metallicity and the quenching boundary from 3200 {LEGA}-{C} galaxies at redshift z{\textasciitilde}0.8},
	shorttitle = {Full spectrum fitting with photometry in ppxf},
	url = {https://ui.adsabs.harvard.edu/abs/2022arXiv220814974C},
	doi = {10.48550/arXiv.2208.14974},
	urldate = {2023-08-02},
	author = {Cappellari, Michele},
	month = aug,
	year = {2022},
	note = {Publication Title: arXiv e-prints
ADS Bibcode: 2022arXiv220814974C},
}

@article{le_exposure_2015,
	title = {Exposure time calculator for {Immersion} {Grating} {Infrared} {Spectrograph}: {IGRINS}},
	volume = {55},
	issn = {0273-1177},
	shorttitle = {Exposure time calculator for {Immersion} {Grating} {Infrared} {Spectrograph}},
	url = {https://ui.adsabs.harvard.edu/abs/2015AdSpR..55.2509L},
	doi = {10.1016/j.asr.2015.03.007},
	urldate = {2023-08-22},
	journal = {Advances in Space Research},
	author = {Le, Huynh Anh N. and Pak, Soojong and Jaffe, Daniel T. and Kaplan, Kyle and Lee, Jae-Joon and Im, Myungshin and Seifahrt, Andreas},
	month = jun,
	year = {2015},
	note = {ADS Bibcode: 2015AdSpR..55.2509L},
	pages = {2509--2518},
}

@article{caglar_llama_2020,
	title = {{LLAMA}: {The} {MBH}-σ⋆ relation of the most luminous local {AGNs}},
	volume = {634},
	issn = {0004-6361},
	shorttitle = {{LLAMA}},
	url = {https://ui.adsabs.harvard.edu/abs/2020A&A...634A.114C},
	doi = {10.1051/0004-6361/201936321},
	urldate = {2023-10-07},
	journal = {Astronomy and Astrophysics},
	author = {Caglar, Turgay and Burtscher, Leonard and Brandl, Bernhard and Brinchmann, Jarle and Davies, Richard I. and Hicks, Erin K. S. and Koss, Michael and Lin, Ming-Yi and Maciejewski, Witold and Müller-Sánchez, Francisco and Riffel, Rogemar A. and Riffel, Rogério and Rosario, David J. and Schartmann, Marc and Schnorr-Müller, Allan and Shimizu, T. Taro and Storchi-Bergmann, Thaisa and Veilleux, Sylvain and Orban de Xivry, Gilles and Bennert, Vardha N.},
	month = feb,
	year = {2020},
	note = {ADS Bibcode: 2020A\&A...634A.114C},
	pages = {A114},
}

@article{schulze_no_2019,
	title = {No signs of star formation being regulated in the most luminous quasars at z ∼ 2 with {ALMA}},
	volume = {488},
	issn = {0035-8711},
	url = {https://doi.org/10.1093/mnras/stz1746},
	doi = {10.1093/mnras/stz1746},
	number = {1},
	urldate = {2023-10-09},
	journal = {Monthly Notices of the Royal Astronomical Society},
	author = {Schulze, Andreas and Silverman, John D and Daddi, Emanuele and Rujopakarn, Wiphu and Liu, Daizhong and Schramm, Malte and Mainieri, Vincenzo and Imanishi, Masatoshi and Hirschmann, Michaela and Jahnke, Knud},
	month = sep,
	year = {2019},
	pages = {1180--1198},
}

@article{ramasawmy_flat_2019,
	title = {A flat trend of star formation rate with {X}-ray luminosity of galaxies hosting {AGN} in the {SCUBA}-2 {Cosmology} {Legacy} {Survey}},
	volume = {486},
	issn = {0035-8711},
	url = {https://doi.org/10.1093/mnras/stz1093},
	doi = {10.1093/mnras/stz1093},
	number = {3},
	urldate = {2023-10-09},
	journal = {Monthly Notices of the Royal Astronomical Society},
	author = {Ramasawmy, Joanna and Stevens, Jason and Martin, Garreth and Geach, James E},
	month = jul,
	year = {2019},
	pages = {4320--4333},
}

@article{stanley_remarkably_2015,
	title = {A remarkably flat relationship between the average star formation rate and {AGN} luminosity for distant {X}-ray {AGN}},
	volume = {453},
	issn = {0035-8711},
	url = {https://doi.org/10.1093/mnras/stv1678},
	doi = {10.1093/mnras/stv1678},
	number = {1},
	urldate = {2023-10-09},
	journal = {Monthly Notices of the Royal Astronomical Society},
	author = {Stanley, F. and Harrison, C. M. and Alexander, D. M. and Swinbank, A. M. and Aird, J. A. and Del Moro, A. and Hickox, R. C. and Mullaney, J. R.},
	month = oct,
	year = {2015},
	pages = {591--604},
}

@article{harrison_no_2012,
	title = {{NO} {CLEAR} {SUBMILLIMETER} {SIGNATURE} {OF} {SUPPRESSED} {STAR} {FORMATION} {AMONG} {X}-{RAY} {LUMINOUS} {ACTIVE} {GALACTIC} {NUCLEI}},
	volume = {760},
	issn = {2041-8205},
	url = {https://dx.doi.org/10.1088/2041-8205/760/1/L15},
	doi = {10.1088/2041-8205/760/1/L15},
	language = {en},
	number = {1},
	urldate = {2023-10-09},
	journal = {The Astrophysical Journal Letters},
	author = {Harrison, C. M. and Alexander, D. M. and Mullaney, J. R. and Altieri, B. and Coia, D. and Charmandaris, V. and Daddi, E. and Dannerbauer, H. and Dasyra, K. and Moro, A. Del and Dickinson, M. and Hickox, R. C. and Ivison, R. J. and Kartaltepe, J. and Floc'h, E. Le and Leiton, R. and Magnelli, B. and Popesso, P. and Rovilos, E. and Rosario, D. and Swinbank, A. M.},
	month = nov,
	year = {2012},
	note = {Publisher: The American Astronomical Society},
	pages = {L15},
}

@inproceedings{herter_performance_2008,
	title = {The performance of {TripleSpec} at {Palomar}},
	volume = {7014},
	url = {https://www.spiedigitallibrary.org/conference-proceedings-of-spie/7014/70140X/The-performance-of-TripleSpec-at-Palomar/10.1117/12.789660.full},
	doi = {10.1117/12.789660},
	urldate = {2023-10-10},
	booktitle = {Ground-based and {Airborne} {Instrumentation} for {Astronomy} {II}},
	publisher = {SPIE},
	author = {Herter, Terry L. and Henderson, Charles P. and Wilson, John C. and Matthews, Keith Y. and Rahmer, Gustavo and Bonati, Marco and Muirhead, Philip S. and Adams, Joseph D. and Lloyd, James P. and Skrutskie, Michael F. and Moon, Dae-Sik and Parshley, Stephen C. and Nelson, Matthew J. and Martinache, Frantz and Gull, George E.},
	month = jul,
	year = {2008},
	pages = {366--373},
}

@ARTICLE{2023ApJ...951....7A,
       author = {{Ayubinia}, Ashraf and {Xue}, Yongquan and {Nguyen Le}, Huynh Anh and {Zou}, Fan and {Wang}, Shu and {He}, Zhicheng and {Kilerci}, Ece},
        title = "{Investigation of Stellar Kinematics and Ionized Gas Outflows in Local (U)LIRGs}",
      journal = {\apj},
         year = 2023,
        month = jul,
       volume = {951},
       number = {1},
          eid = {7},
        pages = {7},
          doi = {10.3847/1538-4357/accf18},
archivePrefix = {arXiv},
       eprint = {2304.09425},
 primaryClass = {astro-ph.GA},
       adsurl = {https://ui.adsabs.harvard.edu/abs/2023ApJ...951....7A}
}

@article{vazdekis_uv-extended_2016,
	title = {{UV}-extended {E}-{MILES} stellar population models: young components in massive early-type galaxies},
	volume = {463},
	issn = {0035-8711, 1365-2966},
	shorttitle = {{UV}-extended {E}-{MILES} stellar population models},
	url = {https://academic.oup.com/mnras/article-lookup/doi/10.1093/mnras/stw2231},
	doi = {10.1093/mnras/stw2231},
	language = {en},
	number = {4},
	urldate = {2023-11-21},
	journal = {Monthly Notices of the Royal Astronomical Society},
	author = {Vazdekis, A. and Koleva, M. and Ricciardelli, E. and Röck, B. and Falcón-Barroso, J.},
	month = dec,
	year = {2016},
	pages = {3409--3436},
}

@article{sexton_bayesian_2020,
	title = {Bayesian {AGN} {Decomposition} {Analysis} for {SDSS} spectra: a correlation analysis of [{O} {\textless}span style="font-variant:small-caps;"{\textgreater}iii{\textless}/span{\textgreater} ] λ5007 outflow kinematics with {AGN} and host galaxy properties},
	volume = {500},
	issn = {0035-8711, 1365-2966},
	shorttitle = {Bayesian {AGN} {Decomposition} {Analysis} for {SDSS} spectra},
	url = {https://academic.oup.com/mnras/article/500/3/2871/5941534},
	doi = {10.1093/mnras/staa3278},
	language = {en},
	number = {3},
	urldate = {2022-10-30},
	journal = {Monthly Notices of the Royal Astronomical Society},
	author = {Sexton, Remington O and Matzko, William and Darden, Nicholas and Canalizo, Gabriela and Gorjian, Varoujan},
	month = nov,
	year = {2020},
	pages = {2871--2895},
}

@ARTICLE{2014MNRAS.443..755H,
       author = {{Husemann}, B. and {Jahnke}, K. and {S{\'a}nchez}, S.~F. and {Wisotzki}, L. and {Nugroho}, D. and {Kupko}, D. and {Schramm}, M.},
        title = "{Integral field spectroscopy of nearby QSOs - I. ENLR size-luminosity relation, ongoing star formation and resolved gas-phase metallicities}",
      journal = {\mnras},
         year = 2014,
        month = sep,
       volume = {443},
       number = {1},
        pages = {755-783},
          doi = {10.1093/mnras/stu1167},
archivePrefix = {arXiv},
       eprint = {1406.4131},
 primaryClass = {astro-ph.GA},
       adsurl = {https://ui.adsabs.harvard.edu/abs/2014MNRAS.443..755H}
}

@ARTICLE{2013A&A...549A..43H,
       author = {{Husemann}, B. and {Wisotzki}, L. and {S{\'a}nchez}, S.~F. and {Jahnke}, K.},
        title = "{The properties of the extended warm ionised gas around low-redshift QSOs and the lack of extended high-velocity outflows}",
      journal = {\aap},
         year = 2013,
        month = jan,
       volume = {549},
          eid = {A43},
        pages = {A43},
          doi = {10.1051/0004-6361/201220076},
archivePrefix = {arXiv},
       eprint = {1210.0566},
 primaryClass = {astro-ph.CO},
       adsurl = {https://ui.adsabs.harvard.edu/abs/2013A&A...549A..43H}
}

@ARTICLE{2017ApJ...836...11G,
       author = {{Gonz{\'a}lez-Alfonso}, E. and {Fischer}, J. and {Spoon}, H.~W.~W. and {Stewart}, K.~P. and {Ashby}, M.~L.~N. and {Veilleux}, S. and {Smith}, H.~A. and {Sturm}, E. and {Farrah}, D. and {Falstad}, N. and {Mel{\'e}ndez}, M. and {Graci{\'a}-Carpio}, J. and {Janssen}, A.~W. and {Lebouteiller}, V.},
        title = "{Molecular Outflows in Local ULIRGs: Energetics from Multitransition OH Analysis}",
      journal = {\apj},
         year = 2017,
        month = feb,
       volume = {836},
       number = {1},
          eid = {11},
        pages = {11},
          doi = {10.3847/1538-4357/836/1/11},
archivePrefix = {arXiv},
       eprint = {1612.08181},
 primaryClass = {astro-ph.GA},
       adsurl = {https://ui.adsabs.harvard.edu/abs/2017ApJ...836...11G}
}

@ARTICLE{2023ApJ...945...59L,
       author = {{Le}, Huynh Anh N. and {Xue}, Yongquan and {Lin}, Xiaozhi and {Wang}, Yijun},
        title = "{[O III] 5007 {\r{A}} Emission Line Width as a Surrogate for {\ensuremath{\sigma}} $_{{\ensuremath{*}}}$ in Type 1 AGNs?}",
      journal = {\apj},
         year = 2023,
        month = mar,
       volume = {945},
       number = {1},
          eid = {59},
        pages = {59},
          doi = {10.3847/1538-4357/acb770},
archivePrefix = {arXiv},
       eprint = {2301.12918},
 primaryClass = {astro-ph.GA},
       adsurl = {https://ui.adsabs.harvard.edu/abs/2023ApJ...945...59L}
}

@ARTICLE{2004ApJ...613..109H,
       author = {{Heckman}, Timothy M. and {Kauffmann}, Guinevere and {Brinchmann}, Jarle and {Charlot}, St{\'e}phane and {Tremonti}, Christy and {White}, Simon D.~M.},
        title = "{Present-Day Growth of Black Holes and Bulges: The Sloan Digital Sky Survey Perspective}",
      journal = {\apj},
         year = 2004,
        month = sep,
       volume = {613},
       number = {1},
        pages = {109-118},
          doi = {10.1086/422872},
archivePrefix = {arXiv},
       eprint = {astro-ph/0406218},
 primaryClass = {astro-ph},
       adsurl = {https://ui.adsabs.harvard.edu/abs/2004ApJ...613..109H}
}

@ARTICLE{1981PASP...93....5B,
       author = {{Baldwin}, J.~A. and {Phillips}, M.~M. and {Terlevich}, R.},
        title = "{Classification parameters for the emission-line spectra of extragalactic objects.}",
      journal = {\pasp},
         year = 1981,
        month = feb,
       volume = {93},
        pages = {5-19},
          doi = {10.1086/130766},
       adsurl = {https://ui.adsabs.harvard.edu/abs/1981PASP...93....5B}
}

@article{le_active_2024,
    title = {Active {Galactic} {Nuclei} and {STaR} {fOrmation} in {Nearby} {Galaxies} ({AGNSTRONG}). {I}. {Sample} and {Strategy}},
    journal = {The Astronomical Journal},
    author = {Le, Huynh Anh N. and Qin, Chen and Xue, Yongquan and KIM, , NGAN N. NGUYEN and Xia, Ruisong and Zhu, Shifu and Lin, Xiaozhi},
    year = {2024},
}

@ARTICLE{2026arXiv260110372Q,
       author = {{Qin}, Chen and {Le}, Huynh Anh N. and {Xue}, Yongquan and {Zhu}, Shifu and {Lin}, Xiaozhi and {Ngan Nhat Nguyen}, Kim},
        title = "{Active Galactic Nuclei and STaR fOrmation in Nearby Galaxies (AGNSTRONG). II: Results for Jetted Type-I AGNs with Strong Ionized Gas Outflows}",
      journal = {arXiv e-prints},
         year = 2026,
        month = jan,
          eid = {arXiv:2601.10372},
        pages = {arXiv:2601.10372},
          doi = {10.48550/arXiv.2601.10372},
archivePrefix = {arXiv},
       eprint = {2601.10372},
 primaryClass = {astro-ph.GA},
       adsurl = {https://ui.adsabs.harvard.edu/abs/2026arXiv260110372Q}
}

@ARTICLE{2015A&A...582A..63C,
       author = {{Cresci}, G. and {Marconi}, A. and {Zibetti}, S. and {Risaliti}, G. and {Carniani}, S. and {Mannucci}, F. and {Gallazzi}, A. and {Maiolino}, R. and {Balmaverde}, B. and {Brusa}, M. and {Capetti}, A. and {Cicone}, C. and {Feruglio}, C. and {Bland-Hawthorn}, J. and {Nagao}, T. and {Oliva}, E. and {Salvato}, M. and {Sani}, E. and {Tozzi}, P. and {Urrutia}, T. and {Venturi}, G.},
        title = "{The MAGNUM survey: positive feedback in the nuclear region of NGC 5643 suggested by MUSE}",
      journal = {\aap},
         year = 2015,
        month = oct,
       volume = {582},
          eid = {A63},
        pages = {A63},
          doi = {10.1051/0004-6361/201526581},
archivePrefix = {arXiv},
       eprint = {1508.04464},
 primaryClass = {astro-ph.GA},
       adsurl = {https://ui.adsabs.harvard.edu/abs/2015A&A...582A..63C}
}

@ARTICLE{2015ApJ...799...82C,
       author = {{Cresci}, G. and {Mainieri}, V. and {Brusa}, M. and {Marconi}, A. and {Perna}, M. and {Mannucci}, F. and {Piconcelli}, E. and {Maiolino}, R. and {Feruglio}, C. and {Fiore}, F. and {Bongiorno}, A. and {Lanzuisi}, G. and {Merloni}, A. and {Schramm}, M. and {Silverman}, J.~D. and {Civano}, F.},
        title = "{Blowin' in the Wind: Both ``Negative'' and ``Positive'' Feedback in an Obscured High-z Quasar}",
      journal = {\apj},
         year = 2015,
        month = jan,
       volume = {799},
       number = {1},
          eid = {82},
        pages = {82},
          doi = {10.1088/0004-637X/799/1/82},
archivePrefix = {arXiv},
       eprint = {1411.4208},
 primaryClass = {astro-ph.GA},
       adsurl = {https://ui.adsabs.harvard.edu/abs/2015ApJ...799...82C}
}

@ARTICLE{2016A&A...591A..28C,
       author = {{Carniani}, S. and {Marconi}, A. and {Maiolino}, R. and {Balmaverde}, B. and {Brusa}, M. and {Cano-D{\'\i}az}, M. and {Cicone}, C. and {Comastri}, A. and {Cresci}, G. and {Fiore}, F. and {Feruglio}, C. and {La Franca}, F. and {Mainieri}, V. and {Mannucci}, F. and {Nagao}, T. and {Netzer}, H. and {Piconcelli}, E. and {Risaliti}, G. and {Schneider}, R. and {Shemmer}, O.},
        title = "{Fast outflows and star formation quenching in quasar host galaxies}",
      journal = {\aap},
         year = 2016,
        month = jun,
       volume = {591},
          eid = {A28},
        pages = {A28},
          doi = {10.1051/0004-6361/201528037},
archivePrefix = {arXiv},
       eprint = {1604.04290},
 primaryClass = {astro-ph.GA},
       adsurl = {https://ui.adsabs.harvard.edu/abs/2016A&A...591A..28C}
}

@ARTICLE{2022MNRAS.512L..54B,
       author = {{Bessiere}, P.~S. and {Ramos Almeida}, C.},
        title = "{Spatially resolved evidence of the impact of quasar-driven outflows on recent star formation: the case of Mrk 34}",
      journal = {\mnras},
         year = 2022,
        month = may,
       volume = {512},
       number = {1},
        pages = {L54-L59},
          doi = {10.1093/mnrasl/slac016},
archivePrefix = {arXiv},
       eprint = {2202.06788},
 primaryClass = {astro-ph.GA},
       adsurl = {https://ui.adsabs.harvard.edu/abs/2022MNRAS.512L..54B}
}

@ARTICLE{2020SciPy-NMeth,
  author  = {Virtanen, Pauli and Gommers, Ralf and Oliphant, Travis E. and
            Haberland, Matt and Reddy, Tyler and Cournapeau, David and
            Burovski, Evgeni and Peterson, Pearu and Weckesser, Warren and
            Bright, Jonathan and {van der Walt}, St{\'e}fan J. and
            Brett, Matthew and Wilson, Joshua and Millman, K. Jarrod and
            Mayorov, Nikolay and Nelson, Andrew R. J. and Jones, Eric and
            Kern, Robert and Larson, Eric and Carey, C J and
            Polat, {\.I}lhan and Feng, Yu and Moore, Eric W. and
            {VanderPlas}, Jake and Laxalde, Denis and Perktold, Josef and
            Cimrman, Robert and Henriksen, Ian and Quintero, E. A. and
            Harris, Charles R. and Archibald, Anne M. and
            Ribeiro, Ant{\^o}nio H. and Pedregosa, Fabian and
            {van Mulbregt}, Paul and {SciPy 1.0 Contributors}},
  title   = {{{SciPy} 1.0: Fundamental Algorithms for Scientific
            Computing in Python}},
  journal = {Nature Methods},
  year    = {2020},
  volume  = {17},
  pages   = {261--272},
  adsurl  = {https://rdcu.be/b08Wh},
  doi     = {10.1038/s41592-019-0686-2},
}

@article{P_rez_Montero_2017,
   title={Ionized Gaseous Nebulae Abundance Determination from the Direct Method},
   volume={129},
   ISSN={1538-3873},
   url={http://dx.doi.org/10.1088/1538-3873/aa5abb},
   DOI={10.1088/1538-3873/aa5abb},
   number={974},
   journal={Publications of the Astronomical Society of the Pacific},
   publisher={IOP Publishing},
   author={Pérez-Montero, Enrique},
   year={2017},
   month=mar, pages={043001} }

@ARTICLE{2020MNRAS.498.4150D,
       author = {{Davies}, R. and {Baron}, D. and {Shimizu}, T. and {Netzer}, H. and {Burtscher}, L. and {de Zeeuw}, P.~T. and {Genzel}, R. and {Hicks}, E.~K.~S. and {Koss}, M. and {Lin}, M.-Y. and {Lutz}, D. and {Maciejewski}, W. and {M{\"u}ller-S{\'a}nchez}, F. and {Orban de Xivry}, G. and {Ricci}, C. and {Riffel}, R. and {Riffel}, R.~A. and {Rosario}, D. and {Schartmann}, M. and {Schnorr-M{\"u}ller}, A. and {Shangguan}, J. and {Sternberg}, A. and {Sturm}, E. and {Storchi-Bergmann}, T. and {Tacconi}, L. and {Veilleux}, S.},
        title = "{Ionized outflows in local luminous AGN: what are the real densities and outflow rates?}",
      journal = {\mnras},
         year = 2020,
        month = nov,
       volume = {498},
       number = {3},
        pages = {4150-4177},
          doi = {10.1093/mnras/staa2413},
archivePrefix = {arXiv},
       eprint = {2003.06153},
 primaryClass = {astro-ph.GA},
       adsurl = {https://ui.adsabs.harvard.edu/abs/2020MNRAS.498.4150D}
}

@ARTICLE{2024A&A...681A..63S,
       author = {{Speranza}, G. and {Ramos Almeida}, C. and {Acosta-Pulido}, J.~A. and {Audibert}, A. and {Holden}, L.~R. and {Tadhunter}, C.~N. and {Lapi}, A. and {Gonz{\'a}lez-Mart{\'\i}n}, O. and {Brusa}, M. and {L{\'o}pez}, I.~E. and {Musiimenta}, B. and {Shankar}, F.},
        title = "{Multiphase characterization of AGN winds in five local type-2 quasars}",
      journal = {\aap},
         year = 2024,
        month = jan,
       volume = {681},
          eid = {A63},
        pages = {A63},
          doi = {10.1051/0004-6361/202347715},
archivePrefix = {arXiv},
       eprint = {2311.10132},
 primaryClass = {astro-ph.GA},
       adsurl = {https://ui.adsabs.harvard.edu/abs/2024A&A...681A..63S}
}

@ARTICLE{2018MNRAS.474..128R,
       author = {{Rose}, Marvin and {Tadhunter}, Clive and {Ramos Almeida}, Cristina and {Rodr{\'\i}guez Zaur{\'\i}n}, Javier and {Santoro}, Francesco and {Spence}, Robert},
        title = "{Quantifying the AGN-driven outflows in ULIRGs (QUADROS) - I: VLT/Xshooter observations of nine nearby objects}",
      journal = {\mnras},
         year = 2018,
        month = feb,
       volume = {474},
       number = {1},
        pages = {128-156},
          doi = {10.1093/mnras/stx2590},
archivePrefix = {arXiv},
       eprint = {1710.06600},
 primaryClass = {astro-ph.GA},
       adsurl = {https://ui.adsabs.harvard.edu/abs/2018MNRAS.474..128R}
}

@ARTICLE{2023MNRAS.520.1848H,
       author = {{Holden}, Luke R. and {Tadhunter}, Clive N. and {Morganti}, Raffaella and {Oosterloo}, Tom},
        title = "{Precise physical conditions for the warm gas outflows in the nearby active galaxy IC 5063}",
      journal = {\mnras},
         year = 2023,
        month = apr,
       volume = {520},
       number = {2},
        pages = {1848-1871},
          doi = {10.1093/mnras/stad123},
archivePrefix = {arXiv},
       eprint = {2301.03999},
 primaryClass = {astro-ph.GA},
       adsurl = {https://ui.adsabs.harvard.edu/abs/2023MNRAS.520.1848H}
}

@ARTICLE{1995MNRAS.272...41S,
       author = {{Storey}, P.~J. and {Hummer}, D.~G.},
        title = "{Recombination line intensities for hydrogenic ions-IV. Total recombination coefficients and machine-readable tables for Z=1 to 8}",
      journal = {\mnras},
         year = 1995,
        month = jan,
       volume = {272},
       number = {1},
        pages = {41-48},
          doi = {10.1093/mnras/272.1.41},
       adsurl = {https://ui.adsabs.harvard.edu/abs/1995MNRAS.272...41S}
}

@BOOK{2006agna.book.....O,
       author = {{Osterbrock}, Donald E. and {Ferland}, Gary J.},
        title = "{Astrophysics of gaseous nebulae and active galactic nuclei}",
         year = 2006,
       adsurl = {https://ui.adsabs.harvard.edu/abs/2006agna.book.....O}
}

@ARTICLE{2023A&A...680A..71H,
       author = {{Hervella Seoane}, K. and {Ramos Almeida}, C. and {Acosta-Pulido}, J.~A. and {Speranza}, G. and {Tadhunter}, C.~N. and {Bessiere}, P.~S.},
        title = "{Investigating the impact of quasar-driven outflows on galaxies at z {\ensuremath{\sim}} 0.3-0.4}",
      journal = {\aap},
         year = 2023,
        month = dec,
       volume = {680},
          eid = {A71},
        pages = {A71},
          doi = {10.1051/0004-6361/202347756},
archivePrefix = {arXiv},
       eprint = {2309.10572},
 primaryClass = {astro-ph.GA},
       adsurl = {https://ui.adsabs.harvard.edu/abs/2023A&A...680A..71H}
}

@ARTICLE{2017A&A...601A.143F,
       author = {{Fiore}, F. and {Feruglio}, C. and {Shankar}, F. and {Bischetti}, M. and {Bongiorno}, A. and {Brusa}, M. and {Carniani}, S. and {Cicone}, C. and {Duras}, F. and {Lamastra}, A. and {Mainieri}, V. and {Marconi}, A. and {Menci}, N. and {Maiolino}, R. and {Piconcelli}, E. and {Vietri}, G. and {Zappacosta}, L.},
        title = "{AGN wind scaling relations and the co-evolution of black holes and galaxies}",
      journal = {\aap},
         year = 2017,
        month = may,
       volume = {601},
          eid = {A143},
        pages = {A143},
          doi = {10.1051/0004-6361/201629478},
archivePrefix = {arXiv},
       eprint = {1702.04507},
 primaryClass = {astro-ph.GA},
       adsurl = {https://ui.adsabs.harvard.edu/abs/2017A&A...601A.143F}
}

@ARTICLE{2022A&A...665A..55S,
       author = {{Speranza}, G. and {Ramos Almeida}, C. and {Acosta-Pulido}, J.~A. and {Riffel}, R.~A. and {Tadhunter}, C. and {Pierce}, J.~C.~S. and {Rodr{\'\i}guez-Ardila}, A. and {Coloma Puga}, M. and {Brusa}, M. and {Musiimenta}, B. and {Alexander}, D.~M. and {Lapi}, A. and {Shankar}, F. and {Villforth}, C.},
        title = "{Warm molecular and ionized gas kinematics in the type-2 quasar J0945+1737}",
      journal = {\aap},
         year = 2022,
        month = sep,
       volume = {665},
          eid = {A55},
        pages = {A55},
          doi = {10.1051/0004-6361/202243585},
archivePrefix = {arXiv},
       eprint = {2206.15347},
 primaryClass = {astro-ph.GA},
       adsurl = {https://ui.adsabs.harvard.edu/abs/2022A&A...665A..55S}
}

@ARTICLE{2022A&A...658A.155R,
       author = {{Ramos Almeida}, C. and {Bischetti}, M. and {Garc{\'\i}a-Burillo}, S. and {Alonso-Herrero}, A. and {Audibert}, A. and {Cicone}, C. and {Feruglio}, C. and {Tadhunter}, C.~N. and {Pierce}, J.~C.~S. and {Pereira-Santaella}, M. and {Bessiere}, P.~S.},
        title = "{The diverse cold molecular gas contents, morphologies, and kinematics of type-2 quasars as seen by ALMA}",
      journal = {\aap},
         year = 2022,
        month = feb,
       volume = {658},
          eid = {A155},
        pages = {A155},
          doi = {10.1051/0004-6361/202141906},
archivePrefix = {arXiv},
       eprint = {2111.13578},
 primaryClass = {astro-ph.GA},
       adsurl = {https://ui.adsabs.harvard.edu/abs/2022A&A...658A.155R}
}

@ARTICLE{1999Ap&SS.266..243C,
       author = {{Calzetti}, D.},
        title = "{UV Emission and bust properties of high redshift galaxies}",
      journal = {\apss},
         year = 1999,
        month = mar,
       volume = {266},
        pages = {243-253},
          doi = {10.1023/A:1002655227201},
archivePrefix = {arXiv},
       eprint = {astro-ph/9902107},
 primaryClass = {astro-ph},
       adsurl = {https://ui.adsabs.harvard.edu/abs/1999Ap&SS.266..243C}
}

@ARTICLE{2002PASP..114..892A,
       author = {{Allington-Smith}, Jeremy and {Murray}, Graham and {Content}, Robert and {Dodsworth}, George and {Davies}, Roger and {Miller}, Bryan W. and {Jorgensen}, Inger and {Hook}, Isobel and {Crampton}, David and {Murowinski}, Richard},
        title = "{Integral Field Spectroscopy with the Gemini Multiobject Spectrograph. I. Design, Construction, and Testing}",
      journal = {\pasp},
         year = 2002,
        month = aug,
       volume = {114},
       number = {798},
        pages = {892-912},
          doi = {10.1086/341712},
       adsurl = {https://ui.adsabs.harvard.edu/abs/2002PASP..114..892A}
}

@ARTICLE{2011ApJ...737...71Z,
       author = {{Zhang}, Kai and {Dong}, Xiao-Bo and {Wang}, Ting-Gui and {Gaskell}, C. Martin},
        title = "{The Blueshifting and Baldwin Effects for the [O III] {\ensuremath{\lambda}}5007 Emission Line in Type 1 Active Galactic Nuclei}",
      journal = {\apj},
         year = 2011,
        month = aug,
       volume = {737},
       number = {2},
          eid = {71},
        pages = {71},
          doi = {10.1088/0004-637X/737/2/71},
archivePrefix = {arXiv},
       eprint = {1105.1094},
 primaryClass = {astro-ph.CO},
       adsurl = {https://ui.adsabs.harvard.edu/abs/2011ApJ...737...71Z}
}
\appendix

\section{Spectra for individual sources}\label{appendixA}
{
The NIR spectra of TPSP for our 6 targets are shown in {Figure}~\ref{fig:nir-spec}. The emission lines that are significant enough are marked. 
ID 2 shows the most and the strongest emission lines with the highest S/N, which can be seen in Table~\ref{tab:lines}.
}

In {Figure}~\ref{fig:opt-spec}, we present the 1D {DBSP (SDSS) spectra for ID 2, ID 4, and ID 5 (ID 1, ID 3, and ID 6){, respectively}}, showcasing the original spectra (black line), the stellar component (red lines), {and the pure emission line} component (blue line), and highlighting some important emission lines (gray dashed vertical lines) present in both the blue and red observed spectra simultaneously. An absorption structure spanning from 6600 to 6900 $\text{\AA}$ in the observed frame is noticeable. Importantly, these structures remain consistent across different redshifts, suggesting that they likely arise from dusty clouds within the Milky Way.

{The electron density $n_e$ and electron temperature $T_e$ are derived from the flux ratios of the forbidden lines. 
We employ the [S II] doublet ratio $F(6716\ \text{\AA})/F(6731\ \text{\AA})$ to diagnose the electron density, following Equation~(16) in \citet{P_rez_Montero_2017}.
Since the [O II] doublet (3726~\AA\ and 3729~\AA) is not resolved in our data, and the [O III] lines have a higher S/N that allows for a more reliable temperature diagnostic, we estimate the electron temperature using the [O III] ratio \((F(4959\ \text{\AA}) + F(5007\ \text{\AA}))/F(4363\ \text{\AA})\), following Equation~(8) in \citet{P_rez_Montero_2017}
The results estimated from SDSS and DBSP spectra are listed in Table~\ref{tab:te_ne_comparison}.
Although trans-auroral lines have been shown to trace denser gas phases and provide more realistic outflow parameters \citep{2018MNRAS.474..128R,2020MNRAS.498.4150D,2023MNRAS.520.1848H,2024A&A...681A..63S}, these lines are too weak to be detected in our current data. 
Thus, we utilize the \hbox{[S II]} doublet to constrain the density.
While this may still underestimate the density, it still provides a more accurate estimate than assuming a fixed value.
In the main text, we use the electron density estimates derived from the SDSS spectra for all six targets to ensure consistency.
}

We compare the normalized model profiles of optical $\rm [O\ III]$  and NIR emission lines in Figure~\ref{fig:vs}. The outflow component of $\rm H_2\ 1\text{-}0\ S(1)$ seems close to the centroid of the profile of $\rm [O\ III]$, indicating the same origin of the outflows.


\renewcommand\thefigure{\Alph{section}\arabic{figure}} 
\setcounter{figure}{0}
\renewcommand{\thetable}{A\arabic{table}}
\setcounter{table}{0}

\begin{table}[htbp]
  \centering
  \caption{Electron Temperature and Density from SDSS and DBSP}
  \label{tab:te_ne_comparison}
  \begin{tabular}{lcccc}
    \hline
    \hline
     & \multicolumn{2}{c}{SDSS} & \multicolumn{2}{c}{DBSP} \\
    ID & $T_e$ & $n_e$ & $T_e$ & $n_e$ \\
    & (K) & (cm$^{-3}$) & (K) & (cm$^{-3}$) \\    \hline
    1 & 19855.70 & 1278.02 & -- & -- \\
    2 & 11751.64 & 431.37 & 10568.85 & 357.03 \\
    3 & 16019.36 & 29.04 & -- & -- \\
    4 & 25506.18 & 82.11 & 21844.95 & 180.53 \\
    5 & 10634.59 & 235.18 & 9649.51 & 363.61 \\
    6 & 20161.93 & 41.51 & -- & -- \\
    \hline
    \hline
  \end{tabular}
\end{table}

\begin{figure*}[t]
    \centering
     \includegraphics[scale=0.85]{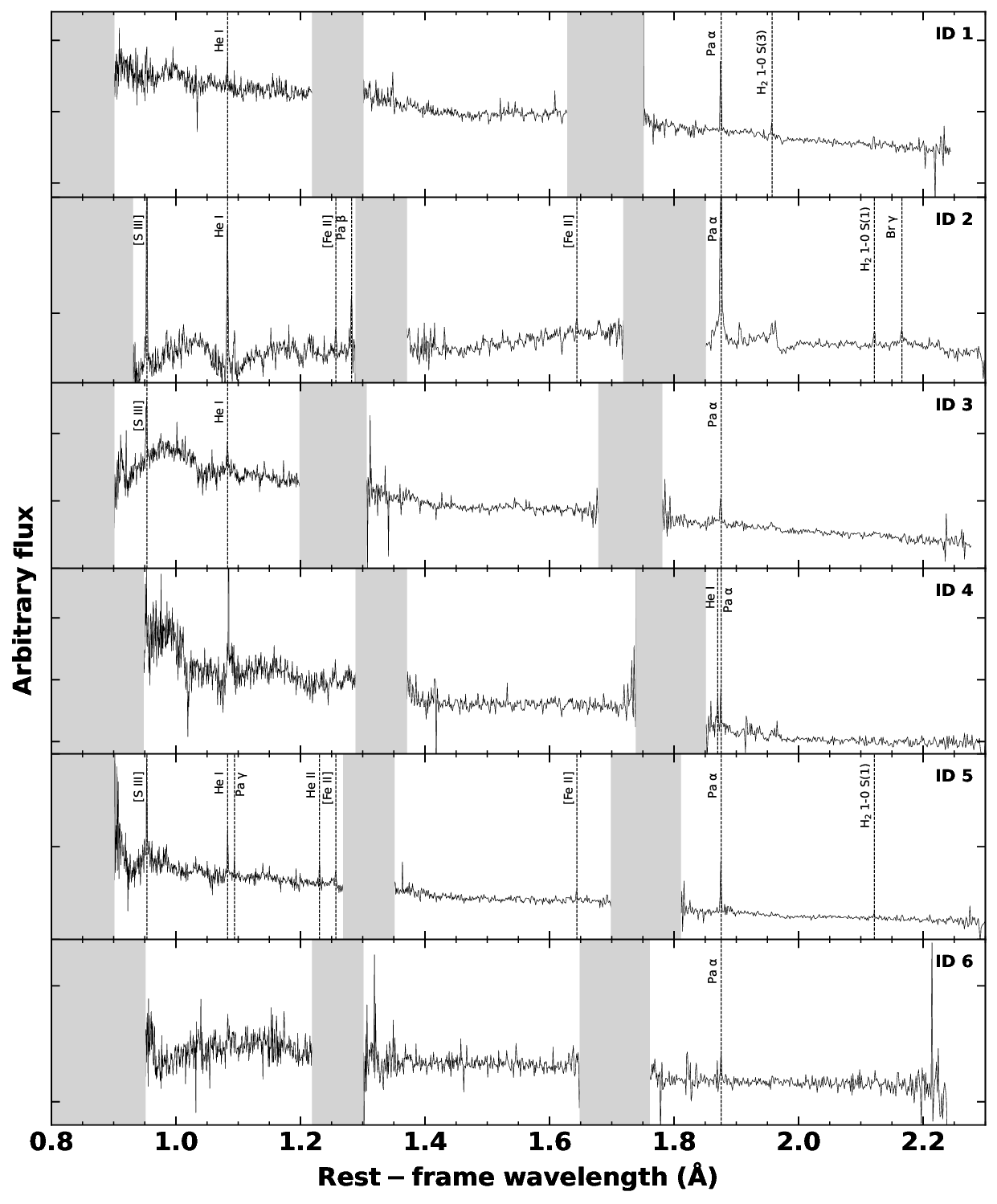}
    \caption{TPSP NIR spectra of the AGNs. The spectra have been corrected by the redshifts concerning the stellar absorption line from the optical spectral analysis of SDSS. The blocked areas indicate the wavelength ranges largely influenced by the sky lines. The emission lines that are significant enough are marked by the dotted lines.}
    \label{fig:nir-spec}
\end{figure*}
\begin{figure*}[t]
    \centering
     \includegraphics[scale=0.85]{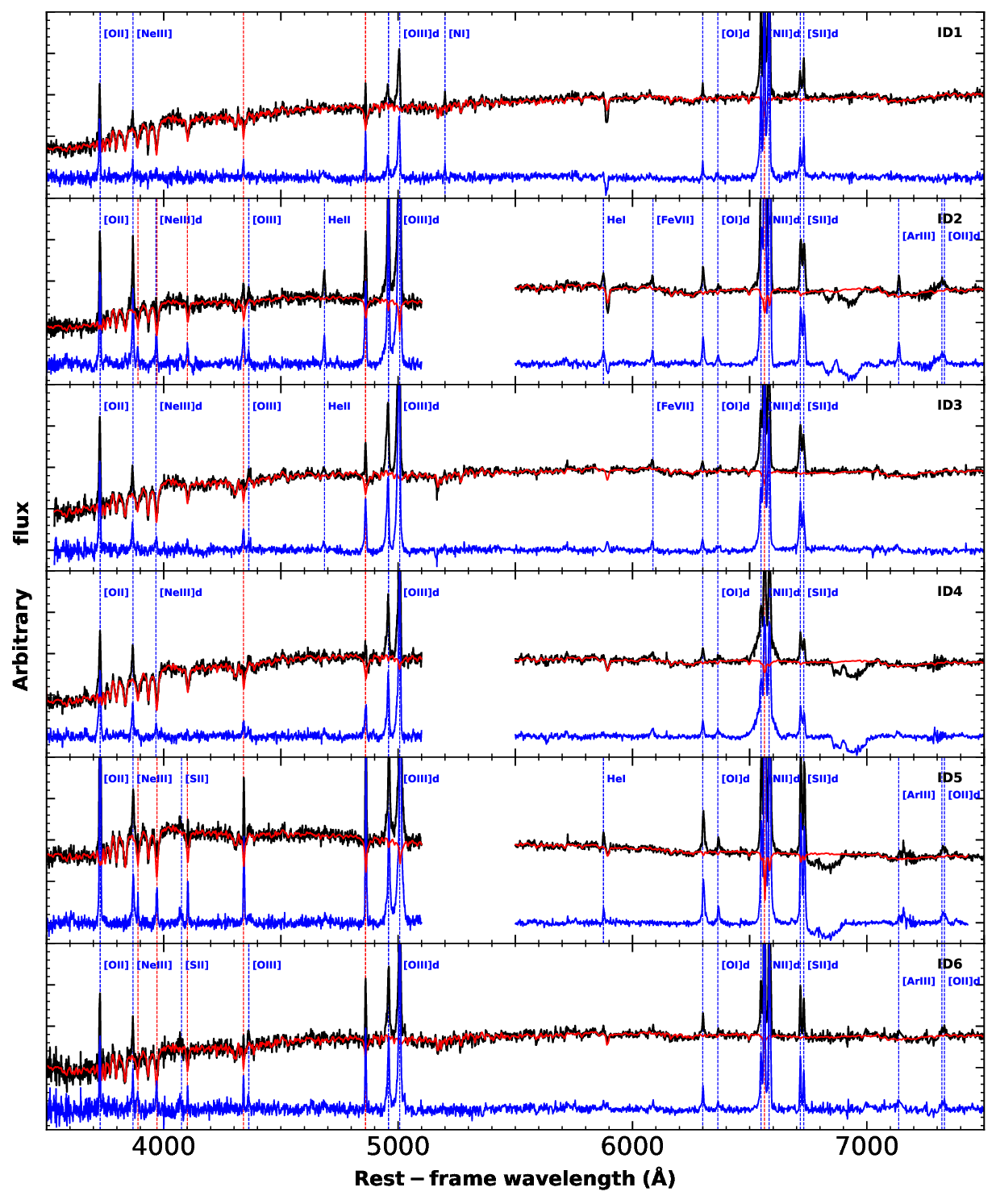}
    \caption{DBSP and SDSS optical spectra of the AGNs (black). The spectra have been corrected concerning the stellar absorption line. The best-fit stellar population models are plotted in red, and the residual emission line spectra are in blue. $\rm H\beta$, $\rm [O\ III]$, $\rm H\alpha + [N\ II]$ and {the} absorption structure range $6600\text{--}6900\ \text{\AA}$ are masked when running pPXF, to make sure {that} a single Gaussian model is reliable to fit {an} emission line. The dashed {lines} indicate the {wavelengths} of some important emission lines in vacuum.}
    \phantomsection
    \label{fig:opt-spec}
\end{figure*}
\begin{figure*}[ht!]
    \centering
    \includegraphics[scale=0.4]{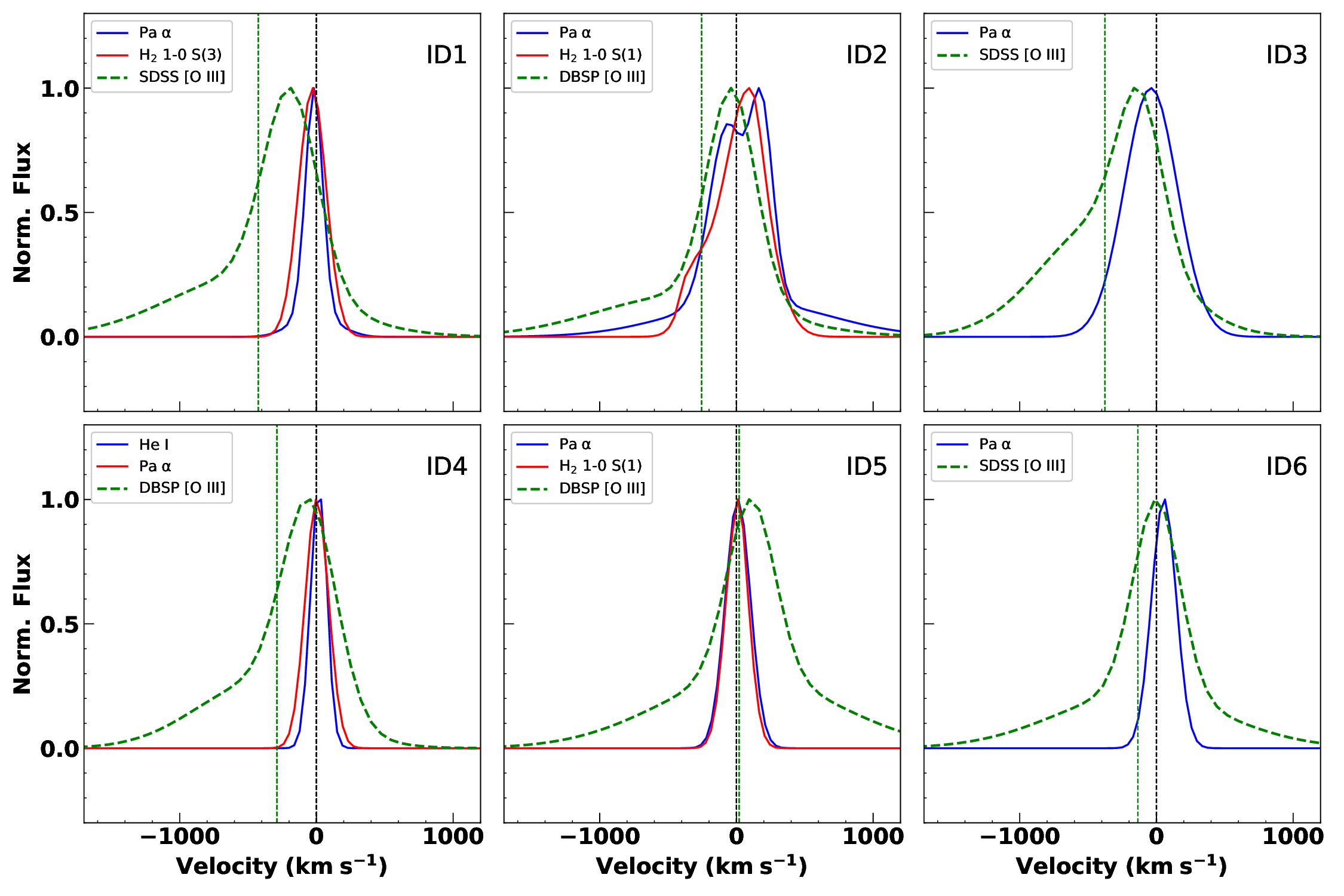}
    \caption{A comparison between optical [O III] line profile (green) and NIR emission lines. The fluxes have been normalized by the peak values of the fitted Gaussian profiles. 
    The dotted lines mark the centroid of the profile fitted by the Gaussian components, and the black {dashed} lines note 0 values.}
    \phantomsection
    \label{fig:vs}
\end{figure*}

\end{document}